\documentclass[iop,apj,appendixfloats]{emulateapj}
\usepackage{apjfonts}
\usepackage{graphicx}
\usepackage{amsmath}
\usepackage{amssymb}

\usepackage{color}

\newcommand{\equ}[1]{eq.~(\ref{eq:#1})}
\newcommand{\equs}[1]{eqs.~(\ref{eq:#1})}
\newcommand{\equss}[1]{(\ref{eq:#1})}
\newcommand{\Equ}[1]{Eq.~(\ref{eq:#1})}

\newcommand{\equnp}[1]{eq.~\ref{eq:#1}}
\newcommand{\equsnp}[1]{eqs.~\ref{eq:#1}}
\newcommand{\equssnp}[1]{\ref{eq:#1}}
\newcommand{\se}[1]{\S\ref{sec:#1}}
\newcommand{\fig}[1]{Fig.~\ref{fig:#1}}
\newcommand{\figs}[1]{Figs.~\ref{fig:#1}}
\newcommand{\figss}[1]{\ref{fig:#1}}
\newcommand{\Fig}[1]{Figure~\ref{fig:#1}}

\newcommand{\tab}[1]{Table~\ref{tab:#1}}
\newcommand{\be}{\begin{equation}}
\newcommand{\ee}{\end{equation}}
\newcommand{\bea}{\begin{eqnarray}}
\newcommand{\eea}{\end{eqnarray}}

\newcommand{\no}{\noindent}

\newcommand{\msun}{M_\odot}

\newcommand{\sy}{\,M_\odot\, {\rm yr}^{-1}}

\newcommand{\ifm}[1]{\relax\ifmmode#1\else$\mathsurround=0pt #1$\fi}
\newcommand{\kms}{\ifmmode\,{\rm km}\,{\rm s}^{-1}\else km$\,$s$^{-1}$\fi}

\newcommand{\Mpc}{\,{\rm Mpc}}

\newcommand{\cmc}{\,{\rm cm}^{-3}}
\newcommand{\cms}{\,{\rm cm}^{-2}}
\newcommand{\kpc}{\,{\rm kpc}}
\newcommand{\pc}{\,{\rm pc}}
\newcommand{\Gyr}{\,{\rm Gyr}}

\newcommand{\Myr}{\,{\rm Myr}}

\newcommand{\K}{\,{\rm K}}

\newcommand{\ltsima}{$\; \buildrel < \over \sim \;$}
\newcommand{\lsim}{\lower.5ex\hbox{\ltsima}}
\newcommand{\gtsima}{$\; \buildrel > \over \sim \;$}
\newcommand{\gsim}{\lower.5ex\hbox{\gtsima}}

\def\omm{\Omega_{\rm m}}

\def\oml{\Omega_{\Lambda}}

\def\s255{S_{\rm 255}}
\def\M10{M_{\rm 10}}
\def\T4{T_{\rm 4}}
\def\z7{(1+z)_{\rm 7}}
\def\L1{\lambda_{\rm s,10}}
\def\fst{f_{\rm s,3}}
\def\t1{\theta_{\rm 0.1}}

\def\Mv{M_{\rm v}}
\def\Rv{R_{\rm v}}
\def\xv{x_{\rm v}}
\def\Vv{V_{\rm v}}

\def\Lj{\lambda_{\rm J}}
\def\Machv{\mathcal{M_{\rm v}}}

\def\Mdots{{\dot{M}_{\rm s}}}
\def\Mdotv{{\dot{M}_{\rm v}}}

\def\rhos{\rho_{\rm s}}

\def\Vs{V_{\rm s}}
\def\Rs{r_{\rm s}}
\def\Ts{T_{\rm s}}
\def\cs{c_{\rm s}}
\def\Ls{\lambda_{\rm s}}

\def\Mls{M_{\rm L,s}}

\def\Mlf{M_{\rm L,Fil}}
\def\Lj{L_{\rm J}}
\def\Mj{M_{\rm J}}

\def\tff{t_{\rm ff}}

\def\fb{f_{\rm b}}

\def\fs{f_{\rm s}}

\lefthead{Mandelker et al.}
\righthead{Globular Clusters and Cold Streams}
\slugcomment{Submitted to ApJ}

\begin{document}
\vspace{1mm}


\title{Cold Filamentary Accretion and the Formation of Metal Poor Globular Clusters and Halo Stars}

%
\author{Nir Mandelker\altaffilmark{1,2,3}, 
Pieter G. van Dokkum\altaffilmark{2}, 
Jean P. Brodie\altaffilmark{4},
Frank C. van den Bosch\altaffilmark{2},
Daniel Ceverino\altaffilmark{5}
\vspace{8pt}}

\altaffiltext{1}
{corresponding author: nir.mandelker@yale.edu}
\altaffiltext{2}
{Department of Astronomy, Yale University, PO Box 208101, New Haven, CT, USA}
\altaffiltext{3}
{Heidelberger Institut f{\"u}r Theoretische Studien, Schloss-Wolfsbrunnenweg 35, 69118 Heidelberg, Germany}
\altaffiltext{4}
{University of California Observatories, 1156 High Street, Santa Cruz, CA 95064, USA}
\altaffiltext{5}
{Universit${\ddot{a}}$t Heidelberg, Zentrum f${\ddot{u}}$r Astronomie, Institut f${\ddot{u}}$r Theoretische Astrophysik, Albert-Ueberle-Str. 2, 69120 Heidelberg, Germany}

\begin{abstract}
We propose that cold filamentary accretion in massive galaxies at high redshift can lead 
to the formation of star-forming clumps in the halos of these galaxies without the presence 
of dark matter sub-structure. In certain cases, these clumps can be the birth places of 
metal poor globular-clusters (MP GCs). Using cosmological simulations, we show that narrow 
streams of dense gas feeding massive galaxies from the cosmic web can fragment, producing 
star-forming clumps. We then derive an analytical model for the properties of streams as a 
function of halo mass and redshift, and assess when these are gravitationally unstable, when 
this can lead to collapse and star-formation in the halo, and when it may result in the formation 
of MP GCs. For stream metalicities $\gsim 0.01 Z_{\odot}$, this is likely to occur at $z>4.5$. 
At $z\sim 6$, the collapsing clouds have masses of $\sim 5-10\times 10^7 \msun$ and the average 
stream pressure is $\sim 10^6 \cmc \K$. The conditions for GC formation are met in the extremely 
turbulent ``eyewall'' at $\sim 0.3\Rv$, where counter-rotating streams can collide, driving very 
large densities. Our scenario can account for the observed kinematics and spatial distribution of 
MP GCs, the correlation between their mass and metalicity, and the mass ratio between the GC system 
and the host halo. For MW mass halos, we infer that $\sim 30\%$ of MP GCs could have formed in this 
way, the remainder likely accreted in mergers. Our predictions for GC formation along circumgalactic 
filaments at high-redshift are testable with JWST.
\end{abstract}


\section{Introduction}
\label{sec:intro}







\smallskip
The origin of globular clusters (GCs) has long challenged models of galaxy formation. 
GCs are bi-modal, with blue, metal-poor (MP), and red, metal-rich (MR), sub-populations 
\citep{Larsen01,Brodie06,Brodie12}. The distribution of metalicities for the two populations 
typically peak at ${\rm [Fe/H]\sim -1.5}$ and $-0.5$ respectively, though both peaks 
tend to shift to higher ${\rm [Fe/H]}$ with increasing galaxy luminosity \citep{Brodie91,
Brodie06,Peng06,Forte09,Kruijssen14}. In the Milky-Way, the populations are typically divided 
at ${\rm [Fe/H]\sim -1.1}$. MP GCs typically comprise $\sim 0.5-0.67$ of the total GC population, 
with the fraction even higher in low mass dwarfs \citep{Strader06,Peng06}. Both populations 
have comparable ages, to within measurement errors, roughly $12.5\Gyr \pm 1\Gyr$ \citep[e.g.][]{Marin09,
vandenburg13,Forbes15}, though most models suggest that MP GCs formed on average earlier than 
MR ones \citep[e.g.][]{Forbes15}. This corresponds to GCs having formed before $z\sim 3$, possibly 
even before the end of reionization. 

\smallskip
The mass functions of both MP and MR GCs are roughly the same. Both have a log-normal distribution, 
with typical masses in the range $\sim 10^4-10^6\msun$ and an average mass of $\sim 2\times 10^5$ 
\citep{Brodie06,Wehner08,Harris13}. The cutoff at the low mass end is thought to be due to disruption 
and evaporation of low mass GCs since their formation, while the initial mass function may have been a 
power law \citep{Elmegreen97,Elmegreen10,Baumgardt98,Fall01,Kravtsov05,Prieto08,McLaughlin08,Kruijssen15}. 
However, others have argued that the tidal forces experienced by GCs in the early Universe 
may have been much weaker than assumed, and thus that it is unlikely that a power law initial mass function 
could have been transformed into a log-normal distribution \citep{Gieles16,Renaud17}. Regardless, the surviving GCs 
should have undergone mass loss since their formation. Models attempting to explain chemical abundance 
anomalies in GCs by invoking multiple stellar populations require the mass at formation to be $>20$ times 
more massive than present day masses (see \citealp{Kruijssen14} for a recent review of such models). However, 
empirical models based on the observational consequences of such massive proto GCs suggest the mass loss ratio 
must be less than 10 \citep{MBK17}, while analytical models of the physics of GC disruption predict mass loss 
ratios as low as $\sim 3$ \citep{Fall01,Kruijssen15}, which is also found in N-body simulations of cluster 
disruption \citep{Webb15}. MP and MR GCs have comparable sizes, with half light radii of $\sim 2-3 \pc$ \citep{Brodie06}.

\smallskip
The MP GCs are particularly enigmatic objects, with some very puzzling properties. 
Their spatial distribution often appears even more extended than the stellar halo 
\citep{Strader11,Forbes12,Durrell14,Kruijssen14}, suggesting that most of them did 
not form inside their host galaxies. The fraction of MP GCs outside of $\sim 10\kpc$ 
is much higher than in the central $\sim 5\kpc$ of galaxies \citep{Harris06}. By contrast, 
MR GCs follow the galaxy light and are associated with the stellar disk and bulge, 
suggesting that they formed along with the thick disk and the bulge, perhaps during 
an unstable clumpy phase \citep{Elmegreen08,Shapiro10,Kruijssen15,Renaud17,M17}, 
or following wet mergers at intermediate redshifts \citep{Ashman92,Kravtsov05,Li14,Li17}. 
MP GCs exhibit a mass-metalicity relation, known as the ``blue tilt", such that more massive 
MP GCs are more metal rich \citep{Strader06,Peng06,Spitler06,Harris06}. This may be due 
to self-enrichment of massive MP GCs \citep{Harris06,Strader08,Bailin09}, but the 
mechanism may depend on environment and does not appear universal \citep{Spitler06,
Strader06,Brodie06}. No such relation is observed for MR GCs \citep{Brodie06,Wehner08}. 
The kinematics of the two populations are also different, with the population of MP GCs 
showing more tangential orbits with significant apparent rotation, as opposed to the mostly 
radial orbits expected if they were mainly accreted \citep{Pota13,Pota15a,Pota15b}. 
The tangential anisotropy of MP GCs increases with distance from the halo center 
\citep{Agnello14}. The MR GCs, on the other hand, have more mixed orbits \citep{Pota15b}.

\smallskip
One of the most puzzling properties of GCs is that the total mass of the GC system 
(GCS) in a galaxy is a near constant fraction of the dark matter halo mass, with a 
ratio of $\sim 2.9\times 10^{-5} \pm 0.28~{\rm dex}$ \citep{Hudson14,Harris15,Harris17}. 
This is in stark contrast to the highly non-linear relation between a galaxy's stellar 
mass and halo mass \citep[e.g.][]{Yang03,Behroozi13}. While a linear relation between 
GCS mass and halo mass exists for the total GC population, it appears mainly driven by 
the MP GCS \citep{Harris15}. This relation holds over 5 orders of magnitude in galaxy 
mass, and in extreme environments, such as entire clusters of galaxies and Ultra-Diffuse 
Galaxies (UDGs) \citep{Harris17,vDokkum17}. Some have suggested that this relation 
is a coincidence, resulting from a stellar mass dependent destruction efficiency for GCs 
combined with the non-linear stellar-to-halo mass relation (\citealp{Kruijssen15}, though 
see \citealp{Fall12} for evidence against a mass dependent destruction efficiency for clusters 
in the local Universe) or as a result of hierarchical galaxy assembly and the central limit 
theorem \citep{EB18}. However, many others have pointed out that this is suggestive of a link 
between GC formation and halo assembly at high redshift \citep[e.g][]{Spitler09,Harris17,MBK17}. 

\smallskip
Recent measurements indicate that the radial extent of GCSs, as measured by their half-number 
radii, is a constant fraction of the halo virial radius, $R_{\rm GC}\sim 0.04\Rv$ \citep{Forbes17}. 
This may be further evidence for an intimate connection betweent the properties of GCSs and 
their dark matter host halos. However, \citet{Hudson17} found a non-linear relation between 
the sizes of GCSs and the virial radii of their host halos, albeit with a smaller sample than 
\citet{Forbes17}. Further observations are needed to clarify this point.

\smallskip
Several classes of models have been envoked to account for the formation of MP GCs. 
Some models propose that they form at the centres of dark matter halos at the earliest stages 
of galaxy formation, prior to reionization \citep{Peebles84,Rosenblatt88,Moore06}. However, 
there is no dynamical evidence for dark matter in GCs \citep{Moore96,Conroy11}. A second class 
of models predict that GCs formed within the gaseous halos surrounding massive galaxies in the 
early Universe, as opposed to in the halo centres, due to instabilities in the halo \citep[e.g.][]{Fall85,
Cen01,Scannapieco04}. A third class of models suggests that MP GCs formed in dwarf galaxies 
in the early Universe, possibly as a result of major mergers \citep{Kravtsov05,Li14,Li17,Kim17}. 
These then merged onto larger galaxies and deposited their GCs in the halos of their new hosts 
\citep{Ashman92,Kravtsov05,Muratov10,Elmegreen12,Kimm16,Kim17,Renaud17}. The similarity between 
properties of GCs and those of young massive clusters (YMCs) in the local Universe has led to the 
suggestion that GCs may be the descendants of ordinary YMC formation at high redshift (\citealp{Elmegreen97,
Kravtsov05,Prieto08,Kruijssen14,Kruijssen15}, though see also \citealp{Renaud17}). \citet{Kruijssen15} 
argues that GC formation is a two stage process, beginning with a rapid-disruption phase in the 
high-pressure environments of high redshift discs until mergers cause them to migrate out into the 
halo, followed by slow evaporation in the halos. While this model is able to reproduce many observed 
properties of GCs and GCSs assuming that all GCs formed at $z\sim 3$, it is not at all clear that all 
GCs formed inside galaxy discs, and other formation mechanisms should be explored \citep[see discussion in][]{Kruijssen14}. 

\smallskip
In particular, the observed connection between GCSs and their dark-matter host halos warrents further 
investigation as to whether such a relation could have existed at their formation. Recently, an empirical 
model has been proposed where MP GCs form directly in their dark matter host halos at $z\gsim 6$ in direct 
proportion to the host halo mass, and then undergo subsequent hierarchical merging of halos and of GCSs 
\citep{MBK17}. It was shown that this can yield $z=0$ GCS masses that are consistent with observations, 
though no physical mechanism was proposed for the formation of GCs in this way.

\smallskip
In this paper, we explore a new formation channel for MP GCs directly in the halos of massive galaxies 
at $z\sim 6$. This is similar in spirit to the second class of models described above, but motivated 
by our new understanding of gas accretion and the structure of the circumgalactic medium (CGM) in massive 
galaxies at high redshift. Such galaxies are predicted to be fed by narrow, dense streams of cold, metal 
poor gas (\se{fil_obs}). We propose that these streams can become gravitationally unstable, leading to 
the formation of massive star-forming clumps in the halos of such galaxies, and in certain cases to the 
formation of MP GCs. 

\smallskip
The remainder of this paper is organised as follows. In \se{fil_obs} we review some of the theoretical 
background and observational evidence for filamentary accretion at high redshift. In \se{fil_3} we use 
a cosmological simulation to illustrate that streams can form bound, star-forming clumps not associated 
with a merging dark-matter halo, and discuss some of their properties. We then discuss recent observations 
that are suggestive of such stream fragmentation. In \se{model} we discuss stream fragmentation analytically. 
We begin by estimating the characteristic radii, densities, and turbulent Mach numbers of streams as a function 
of halo mass and redshift. We then explore whether the streams are gravitationally unstable, and whether they 
can cool and form stars before reaching the central galaxy. Finally, we speculate when this may lead to GC 
formation. In \se{model_disc} we summarize our model and present specific predictions regarding the properties 
of MP GCs and GC systems. We discuss our results and propose avenues for future work in \se{disc}. For the model 
presented in \se{model} as well as the halo mass histories shown in \fig{min_mass}, we adopt cosmological parameters 
$\omm=0.3$, $\oml=0.7$, $H_0=70\kms\Mpc^{-1}$, and a Universal baryon fraction $\fb=0.17$.

\section{Filamentary Accretion - Theory and Observations}
\label{sec:fil_obs}

\smallskip
In this section we review the theoretical background of, and the observational evidence for, 
the existence of cold streams around massive halos, their properties, and how these relate 
to the assumptions of our model. 

\begin{figure}
\begin{center}
\includegraphics[width =0.49 \textwidth]{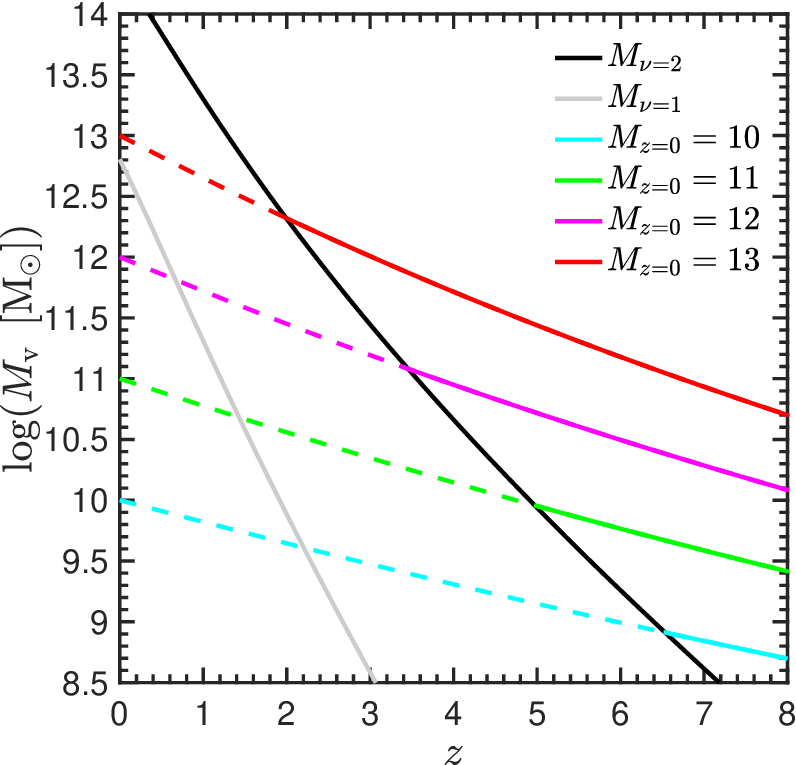}
\end{center}
\caption{Redshift evolution of different halo masses. The black line shows the $2-\sigma$ mass, 
$M_{\rm \nu=2}$, above which halos are likely to be stream fed. The grey line shows the Press-Schechter 
mass, $M_{\rm \nu=1}$, for comparison. Colored lines show the average evolution for halos with $z=0$ 
halo masses ${\rm log}(\Mv/M_{\odot})=10,\,11,\,12$, and $13$. The lines are solid when $\Mv>M_{\rm \nu=2}$ 
and the halos are stream fed, and become dashed when the virial mass drops below $M_{\rm \nu=2}$. 
More massive halos remain stream fed until later times.}
\label{fig:min_mass} 
\end{figure}

\smallskip
The most massive haloes at any epoch lie at the nodes of the cosmic web, and are 
penetrated by cosmic filaments of dark matter \citep[e.g.][]{Bond96,Springel05,
Dekel09}. We refer to such halos as stream fed halos. They represent high-sigma 
peaks, much more massive than the Press-Schechter mass, $M_{\nu=1}$, of typical 
haloes at that time \citep{Press74}. A crude upper limit for the minimal mass of 
a stream fed halo is that of a $2-\sigma$ peak in the cosmic density field, $M_{\rm \nu=2}$ 
\citep{db06,Dekel09}. In \fig{min_mass}, we show both $M_{\rm \nu=1}$ and $M_{\rm \nu=2}$ 
as a function of redshift, computed using \texttt{Colossus} \citep{Diemer15,Diemer15b}. 
We also show the average mass evolution of halos with different $z=0$ masses, computed 
by integrating equation 2 from \citet{Fakhouri10} for the mean mass accretion rate onto 
halos as a function of redshift. Lower mass halos drop below $M_{\rm \nu=2}$ at higher 
redshift, and thus cease being stream fed at earlier times. For example, a $10^{11}\msun$ 
halo at $z=0$ ceases to be stream fed at $z\sim 5$, while a $10^{13}\msun$ halo remains stream 
fed until $z\sim 2$. 

\smallskip
At $z\gsim 2$ the cooling time in sheets and filaments is shorter than the Hubble time \citep{Mo05}. 
Furthermore, in all but the most massive clusters and their progenitors, the gas flowing along these 
dark matter filaments is predicted to be unable to support a stable accretion shock at the filament 
edge \citep{db06,Birnboim16}. Intergalactic gas that accretes onto the filaments from sheets and voids 
remains dense and cold, $T\sim 10^4 \K$, as it free-falls towards the filament axis and settles in 
a narrow dense stream. 
Even in halos above $M_{\rm shock}\sim 10^{12}\msun$, which contain hot gas at the virial 
temperature, the gas in streams is expected to remain cold and penetrate efficiently 
through the hot CGM onto the central galaxy \citep{db06}. 

\smallskip
The above analytical picture is supported by cosmological simulations \citep{Keres05,Ocvirk08,Dekel09,
CDB,FG11,vdv11,Tillson15,Nelson16}. In these simulations, cold streams with widths of a few to ten percent 
of the virial radius penetrate deep into the halo. Many global properties of the streams and their interaction 
with the CGM and with galaxies can be deduced from simulations and compared to observations, as detailed 
below. 

\smallskip
Although cold streams have not been detected directly, there is mounting circumstantial 
observational evidence for their existence. 
Cosmological simulations indicate that the streams maintain roughly constant inflow 
velocities as they travel from the outer halo to the central galaxy \citep{Goerdt15a}. 
The constant velocity, as opposed to the expected gravitational acceleration, 
suggests energy loss into radiation which may be observed as Lyman-$\alpha$ cooling 
emission \citep{Dijkstra09,Goerdt10,FG10}. Radiative transfer models suggest that the 
total luminosity and the spatial structure of the emitted radiation appear similar to 
Lyman-$\alpha$ ``blobs'' observed at $z>2$ \citep{Steidel00,Matsuda06,Matsuda11}. Radiative 
transfer models also show that a central quasar can power the emission by supplying seed 
photons which scatter inelastically within the filaments, producing Lyman-$\alpha$ cooling 
emission that extends to several hundred $\kpc$ and appears similar to observed structures 
\citep{Cantalupo14}. Recent observations using the MUSE integral-field instrument suggest 
that such extended Lyman-$\alpha$ emitting nebulae are ubiquitous around the brightest quasars 
at $z\sim 3.5$ \citep{Borisova16,Vernet17}. The cold streams consist mostly of neutral Hydrogen 
and should also be visible in absorption. They can account for observed 
Lyman-limit systems (LLSs) and damped Lyman-$\alpha$ systems (DLAs) \citep{Fumagalli11,Goerdt12,
vdv12b}. Observations using absorption features along quasar sight-lines to probe the CGM of massive 
SFGs at $z\sim 1-2$ reveal low-metalicity, co-planar, co-rotating accreting material \citep{Bouche13,
Bouche16}, providing further observational support for the cold-stream paradigm. Strong Lyman-$\alpha$ 
absorption has also been detected in the CGM of massive sub-millimeter galaxies (SMGs) at $z\sim 2$ 
\citep{Fu16}. 

\begin{figure*}
\begin{center}
\includegraphics[trim={2.0cm 0 1.7cm 0}, clip, width =0.33 \textwidth]{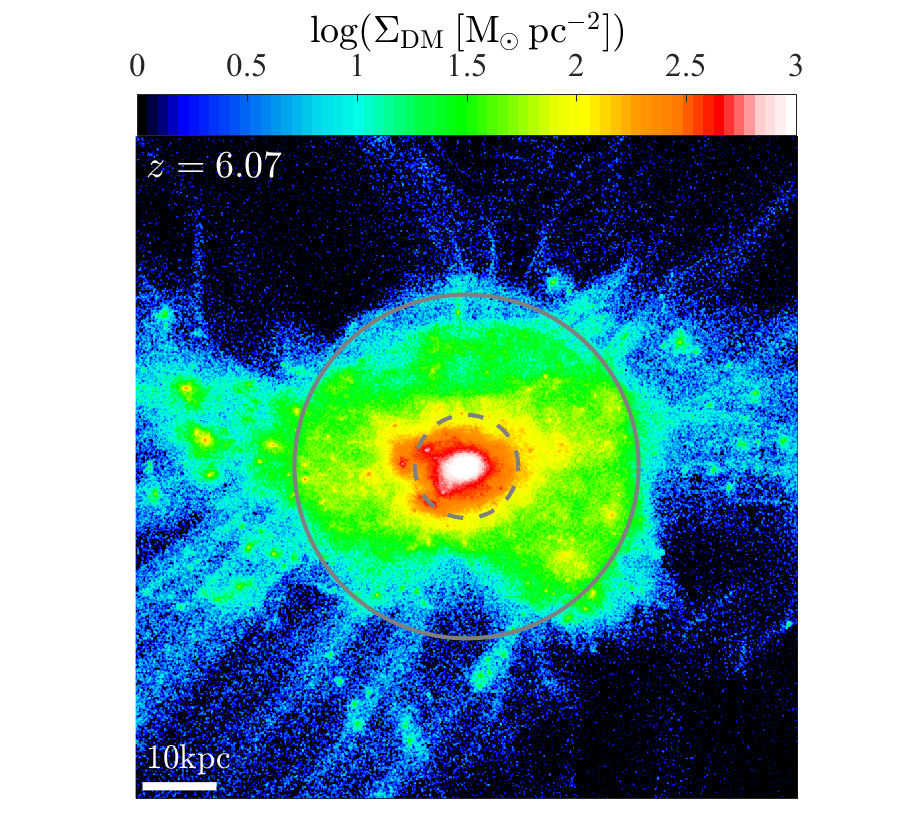}
\includegraphics[trim={2.0cm 0 1.7cm 0}, clip, width =0.33 \textwidth]{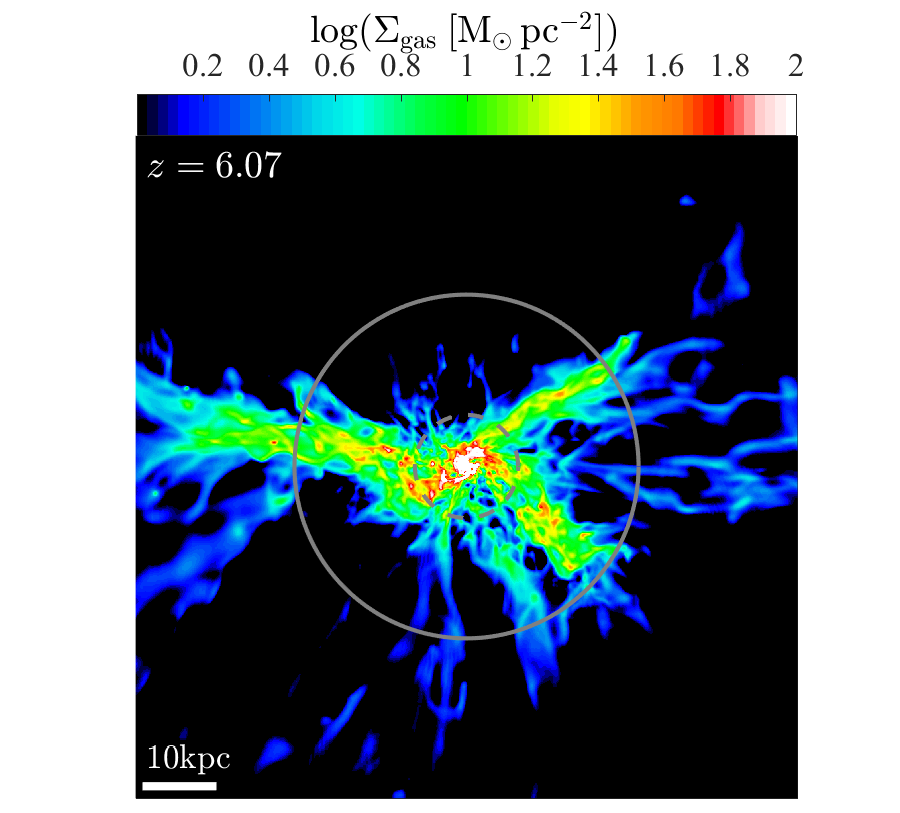}
\includegraphics[trim={2.0cm 0 1.7cm 0}, clip, width =0.33 \textwidth]{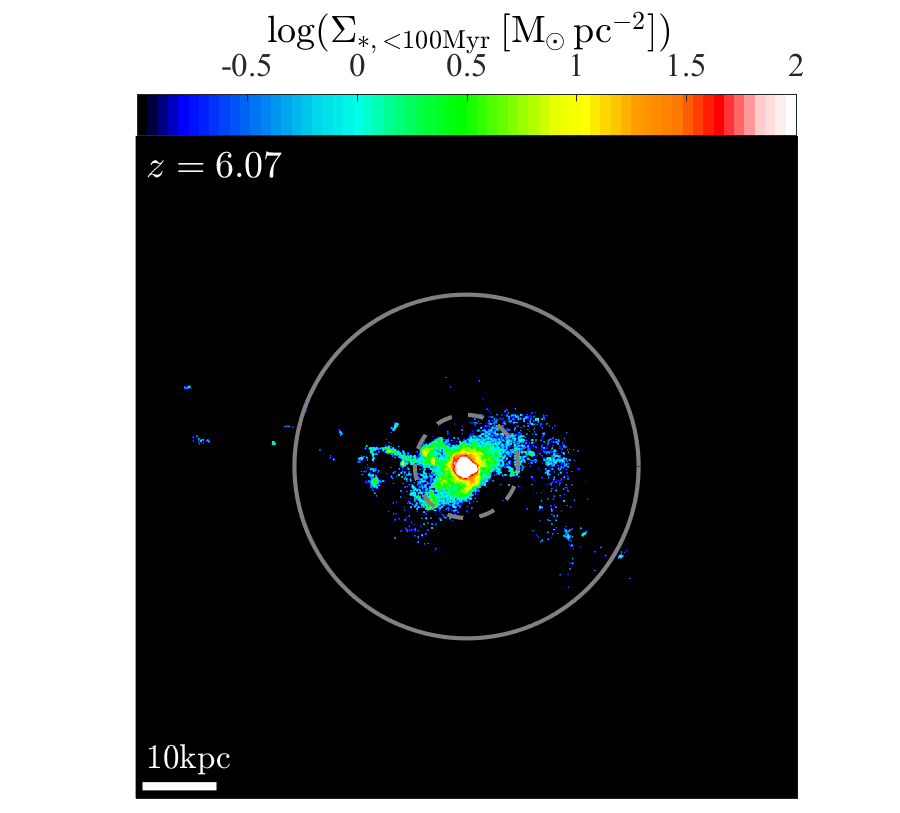}\\
\includegraphics[trim={2.0cm 0 1.7cm 2.5cm}, clip, width =0.33 \textwidth]{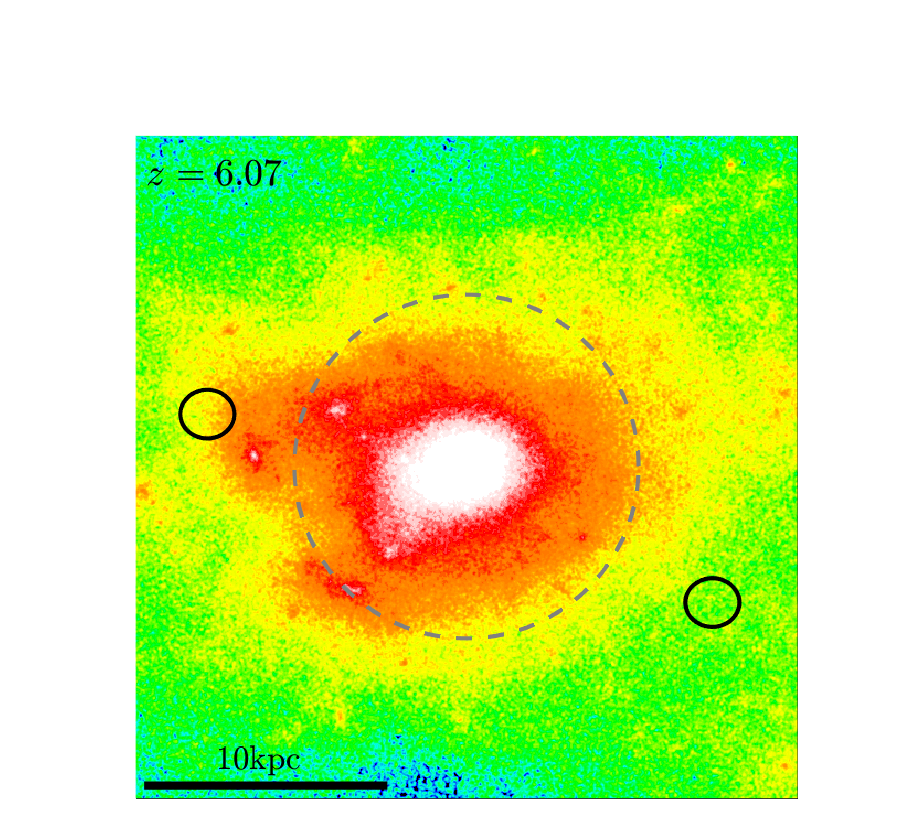}
\includegraphics[trim={2.0cm 0 1.7cm 2.5cm}, clip, width =0.33 \textwidth]{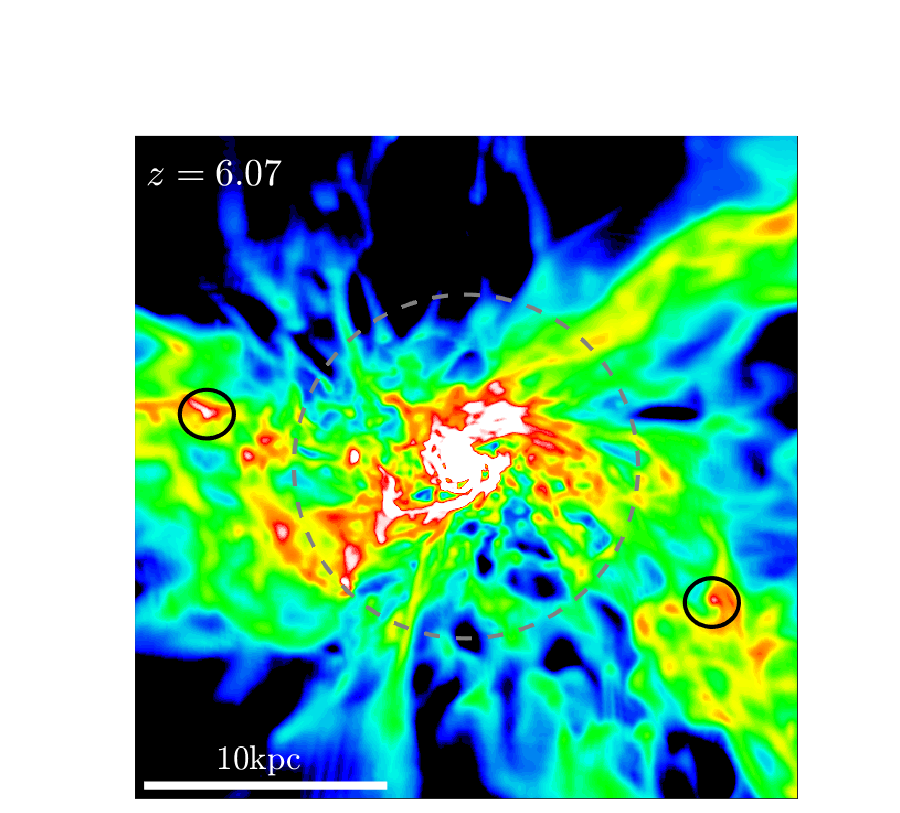}
\includegraphics[trim={2.0cm 0 1.7cm 2.5cm}, clip, width =0.33 \textwidth]{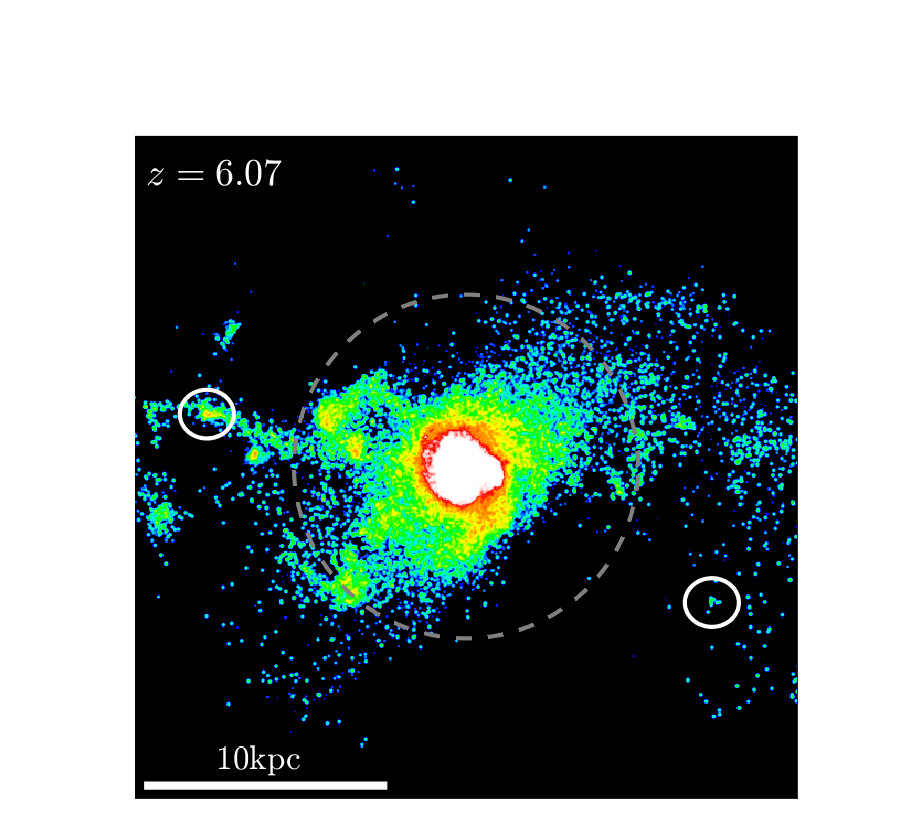}
\end{center}
\caption{Filamentary accretion in a cosmological zoom-in simulation from the \texttt{VELA} simulation suite. 
We show surface densities of dark matter (left), gas (centre), and stars younger than $100\Myr$ (right). The 
top and bottom rows are $100\kpc$ and $30\kpc$ on a side, respectively. Each panel is oriented face on to the 
central disk, and the integration depth of all panels is $40\kpc$. The solid circles in the top row mark the 
halo virial radius, $\Rv$, while the dashed circles in the top and bottom rows mark $0.3\Rv$. The galaxy is 
fed by three dense, narrow gas streams embedded within thicker dark matter filaments. These streams lie in a 
plane that extends beyond $\Rv$, and remain coherent outside of $\sim 0.3\Rv$. In the inner halo, the streams 
can fragment into dense clumps not associated with any dark matter overdensity, where star-formation occurs. 
Two such clumps are circled in the bottom row. Such fragmentation can potentially lead to the formation of MP 
GCs far from the centres of dark matter halos.}
\label{fig:cosmo} 
\end{figure*}

\begin{figure}
\begin{center}
\includegraphics[width =0.49 \textwidth]{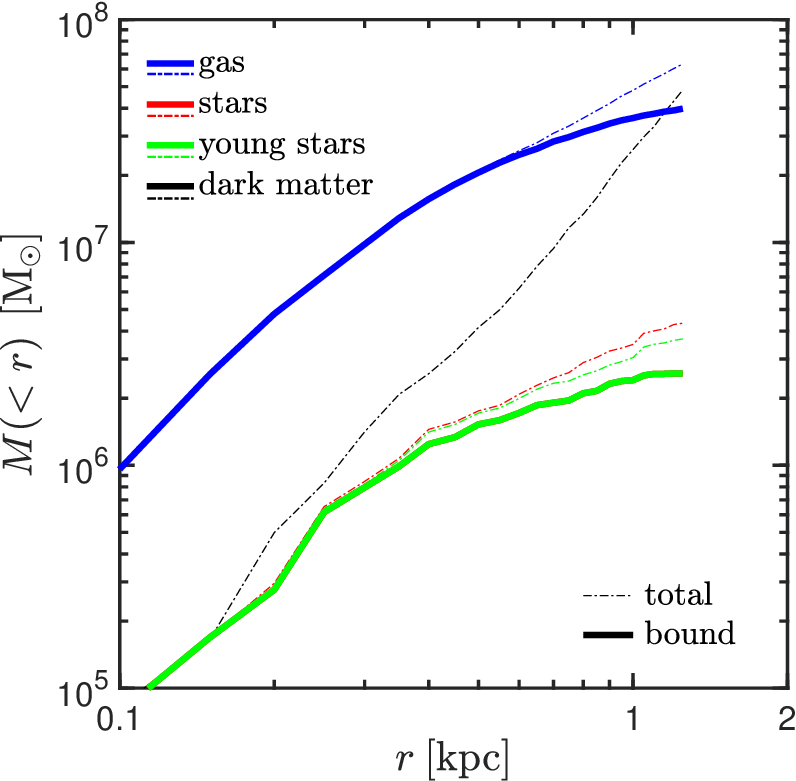}
\end{center}
\caption{Mass profiles for the gas (blue), stars (red), young stars (age$<100\Myr$, green) 
and dark matter (black) in the left-hand clump marked in \fig{cosmo}. The profiles are truncated 
at the tidal radius of the clump, $r_{\rm tid}\sim 1.4\kpc$. Thin dot-dashed lines show the total 
mass profiles while solid lines show the bound mass. Note that all the bound stars are young stars, 
so the solid green line lies on top of the solid red line. No dark matter is bound to the clump, 
while $\gsim 70\%$ of the gas and stars are bound. Within the half mass radius of $r_{\rm h}\sim 450\pc$, 
all of the gas and $\gsim 90\%$ of the stars are bound. The mean Hydrogen number density in the 
inner $100\pc$ is $\sim 7.1\cmc$, while within $r_{\rm h}$ it is $\sim 1.6\cmc$.}
\label{fig:profiles} 
\end{figure}

\smallskip
The streams are predicted to play a key role in the buildup of angular momentum in high-$z$ disk galaxies  \citep{Pichon11,Kimm11,Stewart11,Stewart13,Codis12,Danovich12,Danovich15}. This is due both to vorticity 
within the filaments that spins up dark matter halos \citep{Codis12,Codis15,Laigle15}, and also to an 
impact parameter of the streams with respect to the galaxy centre, typically $\sim 0.3\Rv$ \citep{Keres09,
Danovich15,Tillson15}. Simulations show streams that remain cold and coherent outside of $\sim 0.3\Rv$, 
inside of which a messy interaction region is seen, with strong shocks and highly turbulent flow, where 
the streams collide, fragment, and experience strong torques before spiralling onto the disk in an extended 
ring-like structure \citep{CDB,Danovich15}. Observations of the inner regions of massive, $\Mv>10^{12}\msun$, 
halos at $z\sim 3$ performed with the Cosmic Web Imager (CWI) have revealed extended gaseous structures with 
large angular momentum \citep{Martin14a,Martin14b}. While only a handful of such cases have been observed thus 
far, their structure and kinematics appear very simiar to predictions from cosmological simulations of the 
kinematics of cold streams \citep{Danovich15}. Similar kinematic features have been detected in absorption 
studies of the CGM of massive SFGs at $z\sim 1-2$ \citep{Bouche13,Bouche16}. 

\smallskip
In order to illustrate some of the concepts discussed above, we show in \fig{cosmo} a snapshot of a cosmological 
zoom-in simulation from the \texttt{VELA} simulation suite \citep{Ceverino14,Zolotov15}. The 
simulation was run with the adaptive-mesh refinement (AMR) code \texttt{ART} \citep{Kravtsov97,Kravtsov03,
Ceverino09}. The code incorporates gas and metal cooling, UV-background photoionization with self-shielding in 
gas with Hydrogen number densities $n>0.1\cmc$, stochastic star formation, gas recycling, stellar winds and metal 
enrichment, thermal feedback from supernovae \citep{CDB} and feedback from radiation pressure \citep{Ceverino14}. 
The grid is refined using a quasi-Lagrangian strategy based on the total mass within a cell, up to a maximal 
resolution of $17-35\pc$ (physical) at all times, though the resolution in the outer halo can be significantly 
lower. Details regarding the simulation method and its limitations can be found in \citet{M17}. In \fig{cosmo} 
we show galaxy V19 (see table 1 of \citealp{M17}) at $z=6.07$. At this time, V19 has a virial mass of 
$\sim 1.7\times 10^{11}\msun$, a virial radius of $\sim 26\kpc$, and a stellar mass within $0.1\Rv$ of 
$\sim 7.8\times 10^9\msun$. The last significant merger was a $\sim 1:4$ merger at $z\sim 8$, roughly 
$300\Myr$ before the snapshot shown. 

\smallskip
In \fig{cosmo}, we show maps of the surface density of dark matter (left), gas (centre), and stars younger than 
$100\Myr$ (right). The panels in the top and bottom rows are $100\kpc$ and $30\kpc$ across respectively. 
Concentric solid and dashed circles mark $\Rv$ and $0.3\Rv$ respectively. The integration depth of all panels 
is $40\kpc$, and they are oriented perpendicular to the angular momentum of the central star-forming disk (see 
\citealp{M17} for details on defining this plane). There are three prominent gas streams extending beyond the 
halo virial radius which seem to lie in a plane, the ``stream plane''. While the existence of such a plane is a 
generic feature in cosmological simulations, it does not typically coincide with the plane defined by the disk 
angular momentum as it does in this case \citep{Danovich12}. These gas streams lie at the centres of much wider 
dark matter filaments. The dark matter filaments are difficult to detect within the virial radius, while the 
gas streams remain prominent and coherent until reaching the interaction region at $\sim 0.3\Rv$. 

\section{Fragmentation and Star-Formation in Streams}
\label{sec:fil_3}

\subsection{Stream Fragmentation in Cosmological Simulations}
\label{sec:fil_3a}

\smallskip
Several previous studies have addressed the possibility of clumpy accretion along streams and its 
potential effects on galaxy formation and disc instability \citep[e.g][]{Dekel09,DSC,Genel12b,Goerdt15b}. 
Other studies addressed the possibility of clumps forming due to thermal instabilities in massive halos 
or in streams and its potential effect on heating the CGM in proto-clusters \citep{Dekel08,Birnboim11}. 
However, the formation of baryonic clumps within streams that are not associated with dark matter halos 
has not been studied in cosmological simulations (though \citealp{Pallottini17} do show one such example). 
In \fig{cosmo}, there are several dense clumps in the gas streams in the inner $\sim 0.5\Rv$ that do not appear 
to be associated with dark matter sub-halos. Two such clumps are highlighted with circles in the bottom panels. 
These clumps are forming stars despite not being associated with any dark matter overdensities. 

\smallskip
In order to verify to what extent the highlighted clumps are devoid of dark matter, we 
show in \fig{profiles} the cumulative mass profiles of gas, stars, young stars (age $<100\Myr$), and dark matter 
of the left-hand clump (hereafter clump 1) in \fig{cosmo}. The corresponding profiles for the right-hand clump 
(hereafter clump 2) are qualitatively similar. The profiles are centered on the peak of gas density and extend 
out to the tidal radius of the clump. The tidal radius of a clump falling on a radial trajectory into a massive 
halo is given by the implicit formula \citep{Tormen98} 
\be 
\label{eq:Rtid}
r_{\rm tid} = R_{\rm c}~\left(\frac{m_{\rm c}(r_{\rm tid})}{M_{\rm h}(R_{\rm c})}~\frac{1}{2-d{\rm log} M_{\rm h}(R_{\rm c})/d{\rm log} R}\right)^{1/3},
\ee
{\no}where $m_{\rm c}(r)$ is the total mass of the clump interior to the clumpcentric radius $r$, 
$R_{\rm c}$ is the distance of the clump from the center of the host halo, and $M_{\rm h}(R)$ is 
the total mass of the host halo interior to the radius $R$. The deprojected distance of clump 1 
from the halo center is $R_{\rm c}\simeq 13\kpc \sim 0.5\Rv$, and its tidal radius is 
$r_{\rm tid}\simeq 1.4\kpc$. The total mass profiles interior to $r_{\rm tid}$ are shown 
in \fig{profiles} as thin dot-dashed lines. We then estimate for each gas cell, stellar and dark 
matter particle within $r_{\rm tid}$ whether or not it is bound to the clump, by comparing its 
velocity to the escape velocity from the cell/particle position to the tidal radius, 
\be 
\label{eq:Vesc}
V_{\rm esc}(r) = \left( 2 \int_{r}^{r_{\rm tid}} \frac{Gm_{\rm c}(r)}{r^2} dr \right)^{1/2}.
\ee
The profiles of bound mass are shown in \fig{profiles} as solid lines. There is no dark matter bound to the clump. 
The total mass profile of dark matter roughly scales as $m_{\rm dm}(r)\propto r^3$, indicating a constant background 
of dark matter from the host halo. On the other hand, over $70\%$ of the gas and stellar mass interior to 
$r_{\rm tid}$ is bound. The total bound baryonic mass of the clump is $\sim 5\times 10^7\msun$ and its half 
mass radius is $\sim 450\pc$. Within this radius, all of the gas and over $90\%$ of the stars are bound, and the mean 
density is $\sim 2\cmc$. The mass-weighted mean stellar age of the clump is $\sim 19\Myr$, yielding an average SFR 
of $\gsim 0.2\sy$. 

\begin{figure}
\begin{center}
\includegraphics[trim={2.0cm 0 1.5cm 0}, clip, width =0.49 \textwidth]{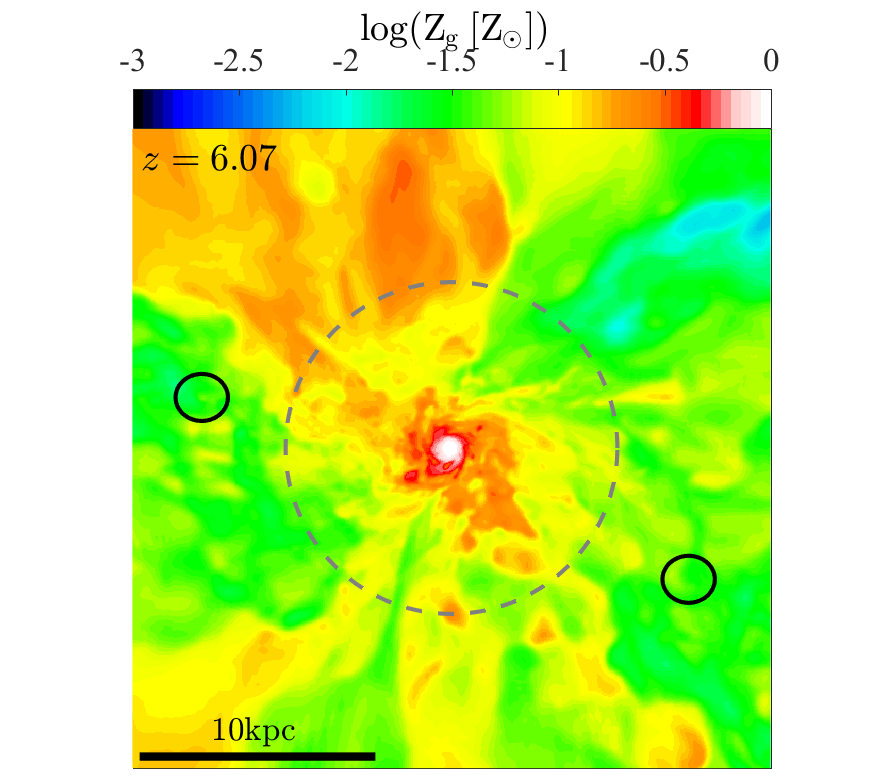}
\end{center}
\caption{Metalicity distribution in the gas streams and central halo. We show the mass-weighted 
average gas metalicity along the line of sight, with the same projection and integration depth as 
in the bottom row of \fig{cosmo}. The gas streams feeding the central galaxy have typical metalicity 
values of $\sim 0.01-0.03~Z_{\odot}$. The clumps forming within the streams have similar metalicity 
values. The high metalicity material outside the streams is low density outflowing gas driven by stellar 
feedback in the central galaxy.}
\label{fig:metals} 
\end{figure}

\smallskip
In \fig{metals} we show the mass-weighted metalicity of the gas phase in the same projection as the bottom 
panels of \fig{cosmo}. Within the cold streams the typical metalicities are ${\rm [Z/H]}\sim -1.5$, consistent 
with observed metalicities of MP GCs. Similarly low metalicities have been found in the simulations analyzed by 
\citet{Fumagalli11,vdv12,Ceverino16}. The mass-weighted mean stellar metalicity within clump 1 is $\sim 0.02 Z_{\odot}$. 

\smallskip
Clump 2 is qualitatively similar to clump 1. It has a tidal radius of $\sim 850\pc$, within which all the dark 
matter is unbound, while $>80\%$ of the gas and stars are bound. Its gas mass is $\sim 2\times 10^7\msun$, 
similar to clump 1, but its stellar mass is only $\sim 2\times 10^5\msun$. Its mean stellar age is $10\Myr$, 
younger than clump 1, with a mean stellar metalicitiy of $\sim 0.025 Z_{\odot}$. 

\smallskip
We stress that we are not claiming that these specific clumps represent MP GCs, which we have no 
hope of resolving in these simulations. In fact these clumps disrupt as they approach the central 
galaxy, depositing some of their stars in the halo and some of them in the disc. However, these 
examples highlight the fact that gas streams may become unstable and form stars and stellar clusters 
far from the centre of any dark matter halo. Beyond the origin of MP GCs, this can have implications 
for the build-up of stellar halos around massive galaxies, and for the frequency of Ly$\alpha$ emitters 
around massive galaxies at high redshift \citep{Farina17}. 

\subsection{Observational Evidence}
\label{sec:fil_2}
\smallskip
There is preliminary observational evidence for the fragmentation of cold streams. Radiative transfer 
models that attempt to reproduce the emission spectra of extended Lyman-$\alpha$ nebulae around luminous 
quasars at $z\sim 3-4$ in halos of mass $\Mv\sim 10^{12}-10^{13}\msun$, indicate that a large amount of 
gas is distributed in dense, compact clumps out to distances of at least $\sim 50\kpc$ \citep{Cantalupo14,
Arrigoni15,Borisova16}. Assuming that the emission is powered mainly by photoionization from the central 
quasar, the models place limits on the clump properties, suggesting densities of $n > 3\cmc$, sizes $R < 20\pc$, 
temperatures $T\sim 10^4{\rm K}$, metalicities $Z<0.1Z_{\odot}$, and typical velocities of $V>500\kms$. 
While such clumps would be unresolved in cosmological simulations, we note that the mean density found in 
clump 1 (\fig{profiles}) is of the same order. Under these conditions, hydrodynamic instabilities induced 
by the motion of these clouds through the hot halo gas should disrupt them on very short timescales 
\citep{Arrigoni15}. This tension may be alleviated if the clumps are not travelling within a hydrostatic 
hot halo, but rather within an inflowing cold stream. It has been argued that such clumps may originate 
from Kelvin Helmholtz Instabilities (KHI) \citep{M16,P18}, Rayleigh Taylor 
Instabilities (RTI) \citep{Keres09}, or thermal instabilities \citep{Cornuault16}. Alternatively, they may 
be due to turbulent fragmentation in the streams (\se{turbulence}). 

\smallskip
Several recent observations highlight the possibility of star-formation occuring inside cold streams. Recent 
MUSE observations of a $z\sim 6.6$ QSO, hosting a $\sim 10^9\msun$ black-hole with an Eddington ratio of 
$\sim 0.7$, have revealed the presence of a Ly$\alpha$ nebula out to distances of $\sim 10\kpc$ from the QSO 
\citep{Farina17}. This is $\sim 25\%$ of the virial radius of a $10^{12}\msun$ halo at $z=6.6$ (\equnp{Rvir}). 
The inferred hydrogen volume density in the nebula is $n_{\rm H}\sim 6\cmc$, similar to the densities in 
streams near $\sim 0.25\Rv$ at $z\sim 6.6$ predicted by our model (\equnp{Rhos}). Furthermore, the authors detected 
a Ly$\alpha$ emitter (LAE) at a projected distance of $12.5\kpc$ from the QSO, with a line-of-sight velocity 
difference of $\sim 560\kms$, comparable to the virial velocity of a $10^{12}\msun$ halo at $z=6.6$ (\equnp{Vvir}), 
located along the direction of the extended nebula. The inferred SFR in this LAE is $\sim 1.3\sy$. Based on the 
QSO-galaxy cross-correlation function, the authors estimate that the probability of finding such a close LAE is 
$<10\%$. However, such a structure is a natural outcome of our model of star-forming clumps in gravitationally 
unstable streams, as detailed below. A similar example has been observed in an extended LAE near a bright 
QSO at $z\sim 3$ \citep{Rauch13}, where it was speculated that feedback from the central QSO triggered a burst 
of star formation at an inferred rate of $\sim 7\sy$ in a nearby dwarf galaxy roughly $17\kpc$ away from the 
QSO. We posit that such bursts of SF may occur directly in the dense, cold gas accreting onto galaxies without 
the need for a satellite dark matter halo or the ``trigger'' from a QSO. 

\smallskip
Finally, ALMA observations of a very massive, $M_*\sim 2\times10^{10}\msun$, galaxy at $z\sim 3.5$ reveal 
a large structure of molecular gas extending out to $\sim 40\kpc$ from the galaxy center \citep{Ginolfi17}. 
This structure has a gas mass of $\sim 2-6\times 10^{11}$, about $60\%$ of which does not appear to be 
associated with either the central galaxy or its satellites. Such a large molecular gas mass cannot be 
accounted for by tidal stripping of satellite galaxies, and kinematic analysis does not reveal coherent 
rotation in the extended structure. This extended structure is also detected in continuum thermal emission, 
with an inferred SFR of $\sim 0.1\sy$. The authors further detect gas rich systems on scales of up to 
$500\kpc$ oriented along the same direction as the extended molecular source, with gas masses in the 
range $10^{10}-10^{11}\msun$ and SFRs of $30-120\sy$. The authors interpret these results as gravitational 
collapse and fragmentation leading to star-formation in the dense inner part of a cold stream feeding the 
central galaxy. A similar finding was reported in the Spiderweb galaxy, a massive proto-cluster at $z\sim 2.2$ 
observed with the Australia Telescope Compact Array (ATCA) and the VLA \citep{Emonts16}. These authors detected 
$\sim 2\times 10^{11}\msun$ of star-forming molecular gas within an extended Ly-$\alpha$ halo up to distances 
of $\sim 100\kpc$ from the central galaxy. 

\section{Analytic Estimates of Stream Fragmentation}
\label{sec:model}

\smallskip
Motivated by the results of the previous section, we present in this section an analytical 
study of gravitational instability and fragmentation in cold streams. In \se{model1} we 
estimate the characteristic sizes and densities of cold streams. In \se{turbulence} we 
identify several sources of turbulence in the streams and estimate the resulting turbulent 
Mach numbers. In \se{model2} we assess the gravitational stability of streams and their 
characteristic fragmentation scales. In \se{model3} we discuss cooling below $10^4\K$ and 
estimate when star-formation can occur in the collapsed clouds. Finally, in \se{GCs} we 
combine all the previous aspects of the model to assess when it may be possible to form 
MP GCs in streams. In the next section, \se{model_disc}, we summarize the main aspects of 
our model and highlight specific predictions for the properties of individual MP GCs and 
GC systems. A schematic illustration of our model, \fig{cartoon}, and a table summarizing 
the main model parameters, \tab{params}, is presented here. Throughout, 
we normalize our results to streams feeding a $10^{10}\msun$ halo at $z=6$, which corresponds 
to the main progenitor of a $\sim 10^{12}\msun$ halo at $z=0$, i.e. a Milky-Way progenitor 
(\fig{min_mass}). However, our model applies to all halos with $\Mv\gsim M_{\rm \nu=2}$, 
and we provide scalings with halo mass and redshift for all derived quantities.  

\subsection{Characteristic Densities and Sizes of Cold Streams}
\label{sec:model1}

\smallskip
We assume that the mass flux along the streams, which have a cylindrical or conical 
shape, is constant along their length until they reach the central halo \citep[e.g.][]{Dekel09}. 
We thus have 
\be 
\label{eq:flux}
\Mdots = \pi \Rs^2 \rhos \Vs = \Mls\Vs.
\ee
{\no}where $\Mdots$ is the mass flux along the stream, $\Rs$ is the stream radius, 
$\rhos$ is its density, $\Vs$ its velocity, and $\Mls=\pi\Rs^2\rhos$ is the \textit{mass 
per unit length} of the stream, hereafter the \textit{line-mass}. As detailed below, we 
allow the stream density and radius to vary with radial position within the halo. 

\smallskip
At $z\gsim 2$, the mean accretion rate of total mass (baryons and dark matter) onto 
the virial radius of a halo with mass $\Mv$ at redshift $z$ is well fit by\footnote{While 
\equ{Mdotv} may slightly underpredict the accretion rate at $z\gsim 5$ \citep{Bosch14}, 
we note that a nearly identical formula can be derived analytically in the EdS regime at 
$z>1$ \citep{Dekel13}.} \citep{Fakhouri10} 
\be 
\label{eq:Mdotv}
\Mdotv \simeq 23 \sy~ \M10^{1.1}\z7^{2.5},
\ee
{\no}where $\M10 = \Mv/10^{10}\msun$ and $\z7=(1+z)/7$. 
The baryonic accretion rate onto the virial radius is given by multiplying \equ{Mdotv} by 
the universal baryon fraction, $\fb$. At the high redshifts we are discussing 
we may assume that the accreted baryons are all gas. Cosmological simulations indicate 
that up to $\sim 50\%$ of the accretion onto the halo is carried by one dominant stream, 
with typically two less prominent filaments carrying up to $\sim 10-20\%$ each \citep{Danovich12}. 
While these simulations focussed on $\Mv \sim 10^{12}\msun$ halos at $z\sim 2$, we assume 
that similar fractions apply in stream fed halos at all redshifts. We adopt $\fs=1/3$ as 
our fiducial value for the fraction of the total accretion carried by a typical stream, and 
obtain for the gas accretion rate along such a stream
\be
\label{eq:Mdot} 
\Mdots \simeq \fs\fb\Mdotv \simeq 1.3\sy~ \fst \M10^{1.1} \z7^{2.5},
\ee
{\no}where $\fst=\fs/(1/3)$ and we have used $\fb=0.17$. Fluctuations in the total accretion rate, 
namely in the normalization of \equ{Mdotv}, can be absorbed into $\fs$ yielding a plausible range 
of $\fst\sim 0.3-3$ \citep{Dekel13}.

\smallskip
The flow velocity along the stream may be written as 
\be 
\label{eq:Vdef}
\Vs = \Machv\Vv,
\ee
{\no}where $\Vv = (G\Mv/\Rv)^{1/2}$ is the virial velocity of the halo, and 
$\Machv$ is an effective ``Mach number", defined as the ratio of the stream 
velocity to the halo virial velocity. Cosmological simulations indicate that the 
streams maintain a roughly constant inflow velocity slightly below the virial value 
as they travel through the halo \citep{Dekel09,Goerdt15a}. We assume $\Machv\sim 1$, 
with an uncertainty of a factor of $\sim 2$. The virial radius\footnote{We define the 
halo virial radius as the radius of a sphere with mean density $18\pi^2\simeq 180$ times 
the mean Universal matter density, valid at $z>1$.} and velocity 
of dark matter halos at $z>1$ are given by \citep[e.g.][]{Dekel13}: 
\be 
\label{eq:Rvir}
\Rv \simeq 10\kpc~\M10^{1/3} \z7^{-1}.
\ee
{\no}and
\be 
\label{eq:Vvir}
\Vv \simeq 66\kms~\M10^{1/3} \z7^{0.5},
\ee

\smallskip
\Equ{flux} can be combined with \equs{Mdot}-\equss{Vvir} to yield the typical line-mass 
of streams feeding a halo with virial mass $\Mv$ at redshift $z$: 
\be 
\label{eq:ML}
\Mls \simeq 1.9\times 10^7 \msun \kpc^{-1}~ \M10^{0.77} \z7^2 \fst \Machv^{-1}.
\ee
{\no}The line-mass of the host dark matter filament is given by dividing 
\equ{ML} by the Universal baryon fraction, 
\be 
\label{eq:MLf}
M_{\rm L,Fil} \simeq 1.1\times 10^8 \msun \kpc^{-1}~ \M10^{0.77} \z7^2 \fst \Machv^{-1}.
\ee

\smallskip
The characteristic radius of dark matter filaments as a function of their line-mass 
and redshift may be estimated by considering the expansion, turnaround and virialization 
of cylindrical top-hat perturbations in an expanding matter dominated 
Universe. This is analogous to the spherical collapse model for dark matter halos. Using 
this model, \citet{FG84} derived the trajectories of collapsing cylindrical shells. They 
found an overdensity of $\sim 3.5$ within the cylinder at turnaround (in the spherical 
collapse model, the overdensity at turnaround is $\sim 5.55$), and that complete radial 
collapse of the shell occurs at roughly $2.5$ times the turnaround time (as opposed to 
twice the turnaround time in the spherical collapse model). However, \citet{FG84} did not 
discuss virialization of the filament. The virial theorem \textit{per unit length} for 
an infinite cylinder is
\be 
\label{eq:virial_cyl}
2T=GM_{\rm L}^2,
\ee
{\no}where $T$ is the kinetic energy per unit length 
and $M_{\rm L}$ is the mass per unit length \citep{Chandrasekhar53,Ostriker64}. Together 
with the expression for the gravitational potential at radius $r$ outside a cylinder with 
line-mass $M_{\rm L}$, 
\be 
\label{eq:cyl_pot_ext}
\Phi(r) = 2GM_{\rm L}{\rm ln}(r/r_0), 
\ee
{\no}where $r_0$ is a reference 
radius, one obtains a ratio of $\sim {\rm exp}(-0.25)\sim 0.78$ between the virial radius 
and the turnaround radius in the cylindrical collapse model (this ratio is $0.5$ in the 
spherical collapse model). This yields a virial overdensity inside the collapsed cylinder 
of\footnote{\citet{Harford11} quote a virial overdensity of $\sim 80$, which is 
what one would infer if one assumes that the ratio of virial radius to turnaround radius 
for collapsed cylinders is $0.5$, as it is for spheres.} $\sim 40$. The mean density in 
a virialized filament is thus 
\be
\label{eq:rho_v_fil} 
\rho_{\rm v,Fil}\simeq 40\Omega_{\rm m}\rho_{\rm crit}(1+z)^3 \simeq 3.8\times 10^{-26}{\rm gr~cm^{-3}}\z7^{3}, 
\ee
{\no}where $\rho_{\rm crit}\sim 9.2\times 10^{-30} {\rm gr~cm^{-3}}$ is the critical density 
of the Universe at $z=0$. Combining this with \equ{MLf}, yields the virial radius of a dark 
matter filament, 
\be 
\label{eq:Rfil}
r_{\rm v,Fil} \simeq 8\kpc~ \M10^{0.38}\z7^{-0.5} \fst^{0.5} \Machv^{-0.5}. 
\ee
{\no}The resulting ratio of filament radius to halo virial radius is 
\be 
\label{eq:Rfil_Rvir}
r_{\rm v,Fil} / \Rv \simeq 0.8~ \M10^{0.05}\z7^{0.5} \fst^{0.5} \Machv^{-0.5}. 
\ee
The relative width of filaments becomes smaller at later times, and has a very weak 
dependence on halo mass. For a $\sim 10^{11}\msun$ halo at $z\sim 6$, this is roughly consistent 
with the most prominent filament seen on the left hand side of \fig{cosmo}. For a 
$10^{12}\msun$ halo at $z=2$, the typical ratio is $\sim 0.66$, which appears consistent 
with high-resolution cosmological zoom-in simulations \citep[e.g.][]{Danovich15,Nelson16}.

\smallskip
The gas streams are significantly narrower than the dark matter filaments, since efficient 
cooling allows the gas to collapse towards the filament axis \citep{db06,Birnboim16}. However, 
the final radius of the gas streams and the mechanism that supports them against gravity has 
not been studied in detail yet. We expect the collapsed stream to be supported by a combination 
of thermal and turbulent pressure (see below), and rotation as evidenced by vorticity in the 
filament \citep{Codis12,Codis15,Laigle15,Birnboim16}. Self-consistently modelling all these 
sources of support is beyond the scope of the current paper, and will require detailed simulations 
of stream evolution. In appendix \se{app}, we show that assuming no rotation in the streams, 
equivalent to assuming that the streams are built by purely radial accretion of gas onto the centers 
of dark matter filaments, yields results that are inconsistent with both cosmological simulations 
and observations. Here, we adopt the opposite extreme and assume that the streams are largely supported 
by rotation, as has been suggested in the literature \citep[e.g.][]{Birnboim16}. This allows us 
to gain a crude estimate of the plausible sizes of streams, which turns out to be much more consistent 
with simulations and observations.

\smallskip
We assume that the stream radius, $\Rs$, scales with the radius of the dark matter filament via a 
contraction factor, $\Ls$, 
\be 
\label{eq:spin}
\Ls \equiv \Rs/r_{\rm v,Fil}.
\ee
{\no}We further assume that the dark matter filament can be approximated as an isothermal cylinder 
truncated at $r_{\rm v,Fil}$. The density profile of an isothermal cylinder is \citep{Ostriker64}:
\be 
\label{eq:iso_fil_dens}
\rho(r) = \rho_{\rm c}\left[1+\left(r/r_{\rm h}\right)^2\right]^{-2},
\ee
{\no}where $\rho_{\rm c}$ is the central density of the filament, and $r_{\rm h}$ is its half mass radius. 
The filament line-mass profile is 
\be 
\label{eq:Mlr_def}
M_{\rm L}(r) = 2\pi \int_{0}^{r} \rho(r)~r~dr.
\ee
{\no}Inserting \equ{iso_fil_dens} yields 
\be 
\label{eq:Mlr}
M_{\rm L}(r) = \Mlf\left[1-\left(1+(r/r_{\rm h})^2\right)^{-1}\right],
\ee
{\no}where $\Mlf=\pi r_{\rm h}^2 \rho_{\rm c}$ is the total line-mass of the filament. The associated circular 
velocity profile is 
\be 
\label{eq:iso_fil_vcirc}
V_{\rm circ}(r) = V_{\infty}\left(r/r_{\rm h}\right)\left[1+\left(r/r_{\rm h}\right)^2\right]^{-1/2},
\ee
{\no}with $V_{\infty}=(2GM_{\rm L})^{1/2}$ the circular velocity at $r\rightarrow \infty$. This can be related 
to the filament virial velocity using \equ{virial_cyl}, yielding $V_{\infty}=\sqrt{2}V_{\rm v,Fil}$. 
If we assume that the specific angular momentum of the gas, $j=\Rs V_{\rm circ}(\Rs)$, is conserved 
during the gas contraction and that it is similar to that of the dark matter in the virialized filament, 
then 
\be 
\label{eq:iso_fil_Ls}
\Ls = \sqrt{2} {\tilde {\lambda}}_{\rm Fil} V_{\rm circ}(r_{\rm v,Fil})/V_{\rm circ}(\Rs),
\ee
{\no}where ${\tilde {\lambda}}_{\rm Fil}=j/(\sqrt{2}r_{\rm v,Fil}V_{\rm v,Fil})$ is the filament 
spin parameter, defined here analogously to the halo spin parameter of \citet{Bullock01}. 
We denote\footnote{$\beta^{-1}$ is analogous to the concentration of spherical dark-matter halos.} 
$r_{\rm v,Fil}/r_{\rm h} = \beta$, and thus $\Rs/r_{\rm h} = \beta\Ls$. If $\beta\Ls<<1$ 
\equ{iso_fil_Ls} reduces to 
\be 
\label{eq:iso_fil_Ls2}
\Ls \simeq \left[2/(1+\beta^2)\right]^{1/4} {\tilde {\lambda}}_{\rm Fil}^{1/2}.
\ee
{\no}This yields $\Ls / {\tilde {\lambda}}_{\rm Fil}^{1/2} \sim 1$, $0.8$ or $0.37$ 
for $\beta=1$, $2$, or $10$ respectively. \Equ{iso_fil_Ls2} relates the stream contraction 
factor to the filament spin parameter. It has been known for some time that the spin 
parameter of dark matter halos has a constant average value of ${\tilde {\lambda}}_{\rm H}\sim 0.035$, 
independent of mass or time \citep{Bullock01}. We assume that the filament spin parameter 
is likewise independent of mass or time, but with a smaller average value since the angular 
momentum of dark matter halos is predicted to originate from the combined angular momentum 
of the filament spin and orbit (impact parameter with respect to the halo centre) \citep{Stewart13,
Danovich15,Laigle15}. For ${\tilde {\lambda}}_{\rm Fil} \sim 0.005-0.02$ and $\beta \sim 1-10$, 
we have $\Ls\sim 0.02-0.14$. We hereafter assume as our fiducial value $\Ls \sim 0.1$, but caution 
that this is highly uncertain and could easily vary by a factor of a few. 

\smallskip
Cosmological simulations indicate that the streams assume a conical shape within the halo due to 
the gravitational attraction towards the halo center \citep[e.g.][]{Dekel09,vdv12}. We thus assume 
$\Rs\propto R$ within the halo, where $R$ is the halocentric radius. Together with \equs{Rfil} and 
\equss{spin}, this yields an expression for the stream radius, 
\be 
\label{eq:Rs}
\Rs(R) \simeq 0.8\kpc~ \xv \M10^{0.38}\z7^{-0.5} \L1 \fst^{0.5} \Machv^{-0.5}, 
\ee 
{\no}where $\L1=\Ls/0.1$ and $\xv={\rm min}(R/\Rv,\,1)$. This is comparable, within a factor $\sim 2$, to 
the size of the large stream seen in \fig{cosmo}. 
We note that in shock-heated halos with $\Mv\gsim 10^{12}$ streams with radii 
$\Rs\lsim 0.05\Rv$ are not expected to reach the central halo, since they will 
be shredded by KHI \citep{M16,P18}. This yields 
a practical lower limit of $\L1\gsim 0.5$ in such massive halos. 

\smallskip
\Equ{Rs} can be used together with \equ{ML} to obtain the average Hydrogen number density in 
the streams\footnote{For a primordial composition of Hydrogen and Helium, the corresponding 
mass density in the streams is roughly $\rhos\simeq 2.3\times 10^{-24}{\rm gr}~n_{\rm H,s}$.}, 
\be 
\label{eq:Rhos}
n_{\rm H,s} \simeq 0.3 \cmc~ \xv^{-2} \z7^3 \L1^{-2},
\ee 
{\no}which is independent of halo mass. 
At $\xv\sim 0.5$, the position of clump 1 in \fig{cosmo}, this yields a density of $\sim 1.2\cmc$, 
comparable to the density within the half mass radius of the clump, where all the gas is still 
bound (\fig{profiles}). When scaled to $z\sim 2$, \equ{Rhos} yields densities that are comparable 
to gas densities found in streams in cosmological simulations, which are in the range $0.001-0.1\cmc$ 
\citep{Goerdt10,FG10,vdv12}. 

\smallskip
Combining \equs{Rs} and \equss{Rhos} we can estimate the column density in a typical stream, 
\be 
\label{eq:Sigs}
\begin{array}{c}
N_{\rm H,s} =2\Rs n_{\rm H,s} \simeq \\
\\
1.4\times 10^{21} \cms~\xv^{-1} \M10^{0.38}\z7^{2.5} \L1^{-1} f_{\rm s,3}^{0.5} \Machv^{-0.5}.
\end{array} 
\ee
{\no}For a $\sim 10^{11}\msun$ halo at $z\sim 6$, this corresponds to a surface mass density of 
$\sim 35\msun\pc^{-2}$, within a factor of 2 of the typical gas surface densities in the streams 
in \fig{cosmo}. Given their high column densities, the streams should be mostly self-shielded 
against the mean UV background radiation at all times. This has been found in simulations at $z\sim 2$ 
\citep{Goerdt10,FG10}.


\subsection{Turbulence in Streams}
\label{sec:turbulence}

\smallskip
In this section we try to estimate the turbulent velocities and turbulent Mach numbers in the streams. 
The temperature of the stream gas is $\Ts \sim 10^4\K$, near the Lyman cooling floor, \citep{db06,Birnboim16}. 
The isothermal sound speed is thus 
\be 
\label{eq:cs}
\cs = \left(\frac{k_{\rm B}\Ts}{\mu}\right)^{1/2} \simeq 8.2\kms~\T4^{0.5},
\ee
{\no}where $\T4=\Ts/10^4 \K$, $k_{\rm B}$ is Boltzmann's constant, and $\mu\sim 1.2m_{\rm p}$ 
is the mean molecular weight, with $m_{\rm p}$ the proton mass. The chosen value for $\mu$ is 
valid for a nearly primordial composition of neutral gas. $\T4$ can vary in the range $\sim 0.8-3$ 
\citep{Goerdt10}, absorbing any possible variation in $\mu$ as well.

\smallskip
The first source of turbulence we consider is accretion of gas onto the streams from the large-scale pancakes 
within which they are embedded \citep{Zeldovich70,Danovich12}, driven by the gravity of the streams and their 
host dark matter filaments. It is well established that such accretion generates turbulence \citep{Klessen10,
Heitsch13,Clarke17,Heigl17}. Based on the models of \citet{Klessen10}, \citet{Heitsch13} predicted the level 
of accretion driven turbulence in cylindrical filaments to be 
\be 
\label{eq:heitsch}
\sigma_{\rm acc} = \left(2\epsilon \Rs V_{\rm acc}^2 \frac{{\dot{M}_{\rm L,s}}}{\Mls}\right)^{1/3},
\ee
{\no}where $\epsilon \sim 0.01-0.1$ is an efficiency parameter which is inversely proprtional to the density 
contrast between the filament and the accreting material \citep{Klessen10}, $V_{\rm acc}$ is the accretion 
velocity onto $\Rs$, and ${\dot{M}}_{\rm L,s}$ is the mass accretion rate onto the stream. 

\smallskip
We estimate the expected radial accretion velocities onto the cold streams by considering the free-fall velocity of 
a cylindrical gas shell starting from rest at the cylindrical turnaround radius, $\sim 1.3 r_{\rm v,Fil}$ (\se{model1}). 
Note that even if the gas has some net rotation velocity, as argued in \se{model1}, we are here only interested in 
the radial component. At $r>r_{\rm v,Fil}$ the gas is accelerated due to a constant line-mass $\Mlf$, which is the 
total line-mass of the filament. The radial velocity at $r_{\rm v,Fil}$ is 
\be 
\label{eq:vrad1}
v_{\rm r}(r_{\rm v,Fil}) = 2\left[G\Mlf {\rm ln}(1.3)\right]^{1/2} \simeq (G\Mlf)^{1/2} = V_{\rm v,Fil},
\ee
{\no}where we have used \equ{cyl_pot_ext} for the gravitational potential outside a filament 
with line-mass $\Mlf$. $V_{\rm v,Fil}=(G\Mlf)^{1/2}$ is the virial velocity of the dark matter 
filament (\se{model1}). At $r_{\rm v,Fil}>r>\Rs$, the line-mass as a function of radius is given 
by \equ{Mlr}. This can be used to compute the change in potential from $r_{\rm v,Fil}$ to $\Rs$ 
and thus the radial velocity at $\Rs$: 
\be 
\label{eq:vrad2}
v^2_{\rm r}(\Rs) = v^2_{\rm r}(r_{\rm v,Fil}) + 2G\Mlf {\rm ln}\left(\frac{1+\beta^2}{1+(\beta\Ls)^2}\right),
\ee
{\no}where we recall that $\beta=r_{\rm v,Fil}/r_{\rm h}$ and $\Ls=\Rs/r_{\rm v,Fil}$. For our fiducial 
value of $\Ls=0.1$ and $\beta=1-10$, we obtain $v_{\rm r}(\Rs) \sim 1.5-3 (G\Mlf)^{1/2} \sim 2.3 (G\Mlf)^{1/2}$. 
Inserting \equ{MLf} into \equ{vrad2} yields 
\be 
\label{eq:accretion}
V_{\rm acc} \simeq 51\kms~ \M10^{0.38} \z7 \fst^{0.5} \Machv^{-0.5}.
\ee
{\no}It is striking that at $z\sim 6$ this is comparable to the virial velocity of the dark matter host halo, 
though it declines more rapidly with redshift. We note that the actual infall velocity onto the streams may be 
smaller that the free-fall velocity computed above if there is some dissipation mechanism acting on the gas as 
it flows towards the filament axis. Since $V_{\rm acc}$ in \equ{accretion} is a factor $\sim 2$ larger than the 
virial velocity of the dark matter filament, this uncertainty should be within a factor of $\lsim 2$ and can be 
absorbed into the efficiency parameter $\epsilon$, discussed below.

\smallskip
The accretion rate onto the stream can be obtained by taking the time derivative of \equ{ML} and 
inserting \equ{Mdotv} and the relation $\z7\simeq (t/0.95~\Gyr)^{-2/3}$. The result is 
\be 
\label{eq:stream_accretion}
\begin{array}{c}
{\dot {M}}_{\rm L,s} \simeq 0.05 \msun \kpc^{-1} {\rm yr}^{-1}~ \M10^{0.77} \z7^{3.5} \fst \Machv^{-1}\\
\\
\cdot \left[0.5+0.63~\M10^{0.1}\z7\right],
\end{array}
\ee
{\no}where the expression in square brackets is $\simeq 1$ for the halo masses and redshifts we are considering. 
Dividing \equ{stream_accretion} by \equ{ML} yields the specific accretion rate of gas onto the stream, 
\be 
\label{eq:specific_accretion}
\frac{{\dot{M}_{\rm L,s}}}{\Mls} \simeq 2.8\Gyr^{-1}~\z7^{1.5}.
\ee

\smallskip
Since $\dot{M}_{\rm L,s}\sim 2\pi \Rs V_{\rm acc} \rho_{\rm acc}$, where $\rho_{\rm acc}$ is the density of 
the accreted gas, we may combine \equs{Rs}, \equss{Rhos}, \equss{accretion}, and \equss{stream_accretion} 
to obtain the density contrast between the accreted gas and the stream gas: 
\be 
\label{eq:density_contrast}
\frac{\rho_{\rm acc}}{\rhos} \simeq 0.02 ~ \xv^3 \L1^3.
\ee
{\no}Inside the halo, the stream gas is $\gsim 100$ times denser than the accreted gas. This appears roughly 
consistent with a visual impression from cosmological simulations \citep[e.g.][figure 7]{Danovich12}. 
Using numerical simulations and analytical arguments, \citet{Klessen10} found that the efficiency of converting 
the inflow kinetic energy to turbulent kinetic energy, $\epsilon$ from \equ{heitsch}, is approximately 
$\epsilon \sim \rho_{\rm acc}/\rhos$, with an uncertainty of a factor of $\sim 3$. We adopt as our fiducial value 
a somewhat conservative $\epsilon=0.01$. Inserting this along with \equs{accretion} and \equss{specific_accretion} 
into \equ{heitsch} yields an estimate for the turbulent velocities driven by accretion 
\be 
\label{eq:accretion_turb}
\sigma_{\rm turb,acc} \simeq 5.2\kms~ \M10^{0.38}\z7 (\epsilon_{1}\,\xv\,\L1)^{1/3}(\fst/\Machv)^{1/2},
\ee
{\no}where $\epsilon_1 = \epsilon/0.01$. Dividing by the sound speed (\equnp{cs}) yields the turbulent Mach number 
associated with accretion 
\be 
\label{eq:accretion_mach}
\mathcal{M}_{\rm turb,acc} \simeq 0.7~ \M10^{0.38}\z7 (\epsilon_{1}\,\xv\,\L1)^{1/3}(\fst/[\Machv\,\T4])^{1/2}.
\ee
{\no}For our fiducial parameters at $z=6$ in the outer halo ($\xv>0.5$), the resulting turbulent Mach numbers 
for halos of mass ${\rm log}(\Mv/\msun)=10$, $11$, and $12$ are $\sim 0.7$, $1.7$, and $4$ respectively. Only 
for the most massive halos is the turbulence highly supersonic. In the inner halo, near $0.3\Rv$, these values 
decrease by a factor of $\sim 0.67$. At $z\sim 2$ the turbulence is transonic for even the most massive halos. 

\smallskip
In addition to accretion, there are several other potential sources of turbulence in the streams. 
As mentioned in \se{fil_obs}, in massive halos with $\Mv\gsim 10^{12}\msun$ that contain hot gas 
at the virial temperature, the streams are susceptible to KHI caused by their interaction with 
the halo gas. The KHI results in oblique shocks within the stream that drive turbulence with 
Mach numbers of order $\mathcal{M_{\rm turb,KHI}}\sim 1$ \citep{P18}. In the 
inner halo, cold streams in hot halos may also be unstable to RTI provided they have an impact 
parameter with respect to the halo centre, placing them above the low density gas in the potential 
well \citep{Keres09}. Finally, cold streams in hot halos may also be thermally unstable \citep{Cornuault16}, 
which can drive highly supersonic turbulence. We note that even in less massive halos, these processes 
may become relevant near the halo centre where hot gas ejected from the central galaxy due to feedback 
may form a hot corona \citep{Sokolowska17}. 

\smallskip
Another source of turbulence and instabilities in the streams is satellite galaxies flowing 
along the streams towards the central galaxy. Simulations suggest that up to $\sim 30\%$ of 
the accretion into massive galaxies at $z\sim 3$ is associated with such satellite galaxies 
in the form of both major and minor mergers \citep{Dekel09}. These galaxies can locally stir 
up the gas in the streams, due both to their gravitational influence on and relative velocity 
with respect to the stream gas, potentially inducing large turbulent motions. Furthermore, winds 
ejected from these satellite galaxies into the stream gas can cause shocks and stir up turbulence, 
and have a profound influence on the structure of streams \citep{FG16}. 

\smallskip
Altogether we estimate that instabilities and feedback can drive turbulence in the streams with 
Mach numbers of $\sim 1$ to a few, preferentially in massive halos with $\Mv>10^{11}\msun$. When 
summed in quadrature to turbulent velocities driven by accretion, the turbulent Mach numbers in 
streams near $\Rv$ may reach $\mathcal{M}_{\rm turb,tot}\sim 1$, $2$ and $4-5$ in halos of mass 
$10^{10}$, $10^{11}$ and $10^{12}\msun$ respectively at $z\sim 6$. However, by $z=3$ these values 
should decrease by $\sim 50\%$.

\subsection{Gravitational Instability and Fragmentation}
\label{sec:model2}

\smallskip
Self-gravitating filaments are unstable to local perturbations at wavelengths larger than 
the 3d Jeans length, provided this is smaller than the radial scale of the filament, even 
in the presence of rotation \citep{Freundlich14}. The Jeans length is given by 
\be 
\label{eq:LJeans_def}
\Lj=[\pi c_{\rm eff}^2/(G\rhos)]^{1/2} ,
\ee
{\no}where $c_{\rm eff}^2 = \cs^2 + \sigma_{\rm turb}^2/3$ represents the combined radial 
support due to thermal and turbulent pressure. Based on the discussion in the previous section, 
we have $c_{\rm eff}/\cs \sim 1.1$, $1.5$, and $2.5$ for streams in halos of mass $10^{10}$, 
$10^{11}$ and $10^{12}\msun$ respectively at $z\sim 6$. At lower redshifts, the contribution 
of turbulent support decreases further. Therefore, in the following discussion we simply adopt 
$c_{\rm eff}=\cs$, and comment where relevant on how this may underestimate the resulting 
fragmentation scales. This has the advantage of examining the degree to which thermal pressure 
alone can support the streams against fragmentation, since streams are often modeled in the 
literature simply as isothermal cylinders at $\Ts\sim 10^4\K$ \citep[e.g.][]{db06,Harford11,M16}.

\smallskip
Using \equs{Rs}, \equss{Rhos} and \equss{cs} we may compute the 
ratio of the thermal Jeans length to the filament diameter, 
\be 
\label{eq:LJeans}
\frac{\Lj}{2\Rs} \simeq 1.4~\M10^{-0.38}\z7^{-1} \fst^{-0.5} \Machv^{0.5}.
\ee
{\no}Note that this is independent of $\Ls$ and of position within the halo. For a 
$10^{10}\msun$ halo, a stream carrying $\fs=0.5$ of the total accretion has a thermal 
Jeans length equal to the stream width. For more massive halos $\Lj<2\Rs$, which means 
that local density perturbations can trigger gravitational collapse within the streams. 
It can easily be verified that adopting the turbulent Jeans length does not change this 
conclusion. 
For streams where $\Lj<2\Rs$, the characteristic mass of gravitationally unstable clumps 
is given by the Jeans mass, 
\be 
\label{eq:Mclump}
\Mj \sim (4\pi/3)\rhos (\Lj/2)^3 \simeq 5.85\times 10^7 \msun~ \xv \z7^{-1.5} \L1 . 
\ee 
{\no}Note that the thermal Jeans mass is independent of halo mass, and scales linearly 
with halocentric radius. At $z=6$, the turbulent Jeans mass is larger by a factor of 
$\sim 3$ and $\sim 15$ in streams feeding halos of mass $10^{11}$ and $10^{12}\msun$. 
At lower redshifts the difference between the thermal and turbulent Jeans masses becomes 
smaller. In appendix \se{app2} we show that the effects of the tidal fields induced by 
the host halo and the host dark-matter filament on the fragmentation scale are both negligible.

\smallskip
In order to ascertain whether such a Jeans unstable cloud will have time to collapse 
before the stream reaches the central galaxy, we compare the free fall time, 
$\tff=[3\pi/(32G\rhos)]^{1/2}$, to the inflow time from a given radius, $t_{\rm inflow}=R/\Vs$. 
We find 
\be 
\label{eq:tff}
\frac{\tff}{t_{\rm inflow}} \simeq 0.6~ \L1 \Machv.
\ee
{\no}This is independent of halo mass, redshift, or position within the halo. 
\textit{Wherever the perturbation is seeded, it will collapse in roughly half 
the time it takes to reach the central galaxy.} 

\smallskip
For long wavelength perturbations, larger than the filament diameter, the above local 
stability criterion cannot be used, and we must instead examine the global stability 
of the filament. A self-gravitating isothermal filament is unstable to global axisymmetric 
perturbations if its line-mass is larger than a critial value $M_{\rm L,crit}=2\cs^2/G$ 
\citep[e.g.][]{Ostriker64,Inutsuka92}. This can be thought of as a filamentary ``Jeans line-mass'', 
above which thermal pressure cannot prevent global radial collapse of the filament. The ratio of the 
line-mass in a typical stream to this critical value is 
\be 
\label{eq:Mlcrit}
\frac{\Mls}{M_{\rm L,crit}} \simeq 0.6~ \M10^{0.77} \z7^2 \fst \Machv^{-1} \T4^{-1}.
\ee
{\no}At $z=6$, streams feeding halos more massive than $\sim 2\times 10^{10}\msun$ are supercritical, 
and thus globally unstable. At $z=3$, the critical halo mass for unstable streams is $\sim 2\times 10^{11}\msun$, 
smaller than $M_{\rm \nu=2}$ at that redshift (\fig{min_mass}). The characteristic collapse time for a 
supercritical filament is $t_{\rm coll}\sim 3(4\pi G \rhos)^{-1/2}$ \citep{Inutsuka92,Heitsch13}. 
Comparing this to the inflow time we find 
\be 
\label{eq:tcoll}
\frac{t_{\rm coll}}{t_{\rm inflow}} \simeq 0.9~ \L1 \Machv,
\ee
{\no}so the stream has time to globally fragment before reaching the central galaxy. This global instability 
results in the formation of dense cores separated by a few times the filament diameter \citep{Inutsuka92,
Clarke16}. These cores then proceed to fragment on the local Jeans scale discussed above \citep{Clarke16,Clarke17}. 

\smallskip
Our analysis thus suggests that most streams are supercritical with fragmentation times shorter than 
the halo crossing time. This is exacerbated further by the additional inwards gravitational force of 
the dark matter filament. This means that there must be some additional source of support in the streams 
which prevents catastrophic fragmentation in cosmological simulations. This could be rotation (\se{model1}), 
turbulence (\se{turbulence}), or artificial support caused by low resolution. However, \citet{Heitsch13} 
found that accretion-driven turbulence, which we expect to be the dominant source of accretion in most cases, 
can slow down the global collapse of supercritical filaments but it cannot halt it. The basic reason is that 
the line-mass of the stream grows faster than the resulting turbulent support, and therefore the stream remains 
super-critical. This has been confirmed by \citet{Clarke17}, who simulated a self-gravitating isothermal filament 
growing by accretion from its surroundings. They found that as long as the accretion flow itself was not highly 
turbulent, gravitational collapse and fragmentation occurs in two stages, first on large scales set by filamentary 
fragmentation, and then on small scales set by the local Jeans scale.

\smallskip
To summarize, we find that most streams feeding massive halos should be gravitationally unstable to both short 
and long wavelength perturbations. The former result in direct three dimensional collapse on the Jeans scale, 
of clouds with masses given by \equ{Mclump}. The latter result first in two dimensional filamentary collapse, 
followed by subsequent three dimensional collapse on the Jeans scale, leading to collapsed clouds with the same 
mass as in the former case. Both of these processes can act in less than a virial crossing time, particularly if 
the stream is relatively narrow with $\L1\sim 0.5$. 

\subsection{Cooling and Star Formation}
\label{sec:model3}

\smallskip
In order for star-formation to occur in the collapsing gas clumps, they must be able to cool from the initial 
stream temperatures of $\Ts\sim 10^4\K$ down to $\sim 10\K$. If the streams are indeed self shielded against the 
UV background then at $10^4\K$ they are mostly neutral, as assumed above. At metalicities of $Z>10^{-3}Z_{\odot}$, 
and at the characteristic densities and temperatures of the streams, the dominant cooling process is emission in 
the ${\rm [C\,II]\:158 \mu m}$ line \citep{Krumholz12,Pallottini17}. For a cloud of gas with a mean Hydrogen number 
density ${\bar n}_{\rm H}=n_0 \cmc$, metalicity $Z=0.01Z_{\odot}~Z_2$, temperature $T=10^4\K~\T4$, and clumping factor 
$\mathcal{C}=<n_{\rm H}^2>/{\bar n}_{\rm H}^2=10\mathcal{C}_{10}$, the ratio of the cooling time to the free-fall time 
of the cloud is \citep[][equation 6]{Krumholz12} 
\be 
\label{eq:cool}
\frac{t_{\rm cool}}{\tff} \simeq 9.1 {\rm exp}(0.009 / \T4) ~ Z_2^{-1} \mathcal{C}_{10}^{-1} n_0^{-1/2} \T4.
\ee
{\no}For a collapsing cloud, this ratio decreases as the density increases during the collapse. However, 
since both the free-fall time and the cooling time are dominated by their initial stages near the onset 
of collapse, this initial ratio is representative of the final ratio. Inserting \equ{Rhos} for the mean 
density in the streams and using $\T4\sim 1$ this yields 
\be 
\label{eq:cool2}
\frac{t_{\rm cool}}{\tff} \simeq 16.8 ~ Z_2^{-1} \mathcal{C}_{10}^{-1} \T4 ~ \xv \z7^{-1.5} \L1.
\ee
{\no}Note that this depends on position in the halo through the stream density. 

\smallskip
Multiplying \equ{cool2} by \equ{tff} yields the ratio of cooling time in the streams to the inflow time, 
\be 
\label{eq:cool3}
\frac{t_{\rm cool}}{t_{\rm inflow}} \simeq 10.1 ~ Z_2^{-1} \mathcal{C}_{10}^{-1} \T4 ~ \xv \z7^{-1.5} \L1^2 \Machv.
\ee
{\no}This ratio increases towards lower redshifts, as the streams become less dense. We define the cooling 
redshift, $z_{\rm cool}$ as the redshift where the cooling time is equal to the inflow time, 
\be 
\label{eq:zcool}
1+z_{\rm cool} \simeq 32.7 ~ \left(Z_2 \mathcal{C}_{10}\right)^{-2/3} \left(\T4 \Machv \L1^2 \right)^{2/3} \xv^{2/3}.
\ee
{\no}At $z<z_{\rm cool}$, the cooling time in the collapsing clump is longer than the inflow time to the halo center, 
and we do not expect much star-formation in the clump. However, at $z>z_{\rm cool}$, the clump may experience a burst 
of star-formation before reaching the central galaxy.

\smallskip
Clumping factors of $\mathcal{C}\sim 5-10$ are not unreasonable at $z\sim 6$ given the levels of turbulence and 
substructure in the streams. In fact observations suggest that the clumping factors may be even higher in very 
massive halos (\se{fil_2}). We therefore assume $\mathcal{C}_{10}\gsim 0.5$ at $z\sim 6$, possibly approaching 
$\mathcal{C}_{10}\sim 1$ for $\sim 10^{12}\msun$ halos where the turbulent velocities are very large. However, 
as the turbulent velocities decrease towards lower redshift the associated clumping factors may decrease as well. 
For a relatively narrow stream with $\L1=0.5$ near the inner halo at $\xv=0.3$, with a metalicity of $Z_2=2$ 
(\fig{metals}) and a clumping factor of $\mathcal{C}_{10}=0.5$, this yields $z_{\rm cool} \sim 4.8$. At $z=6$, this 
yields $t_{\rm cool}\sim 2.1\tff \sim 0.6 t_{\rm inflow}$. In this case, the clump can cool and proceed to star-formation 
on a free-fall timescale, before reaching the central galaxy. Furthermore, since the initial cooling timescale is 
longer than the free-fall timescale of the cloud, we expect relatively large contraction factors for these clouds 
before the onset of star-formation. This can increase the mean densities in the clouds by considerable amounts 
compared to the mean densities in the streams given by \equ{Rhos}. It is also worth noting that such a cloud can cool 
and form stars \textit{before} forming any appreciable fraction of ${\rm H}_2$ \citep{Krumholz12}.

\smallskip
In order to obtain $z_{\rm cool} \sim 2$ in a stream with fiducial parameters, we require either larger metalicities 
closer to $\sim 0.1 Z_{\odot}$, or larger clumping factors. Overall, we predict star-formation to be less likely to 
occur in the streams outside the central galaxy at $z \lsim 4$, unless the metalicities are $\sim 0.05-0.1Z_{\odot}$, 
larger than indicated by most cosmological simulations \citep{Fumagalli11,vdv12,Ceverino16}. 

\smallskip
In order for collapse to occur the clump must also dissipate its turbulent support. The initial clump radius is of 
order the stream radius (\equnp{LJeans}). When accounting for turbulence in the Jeans length, this remains true even 
at $\Mv\sim 10^{12}\msun$. The dissipation rate of turbulence over a length scale $\Rs$ is equal to 
$t_{\rm diss} \sim \Rs/\sigma_{\rm turb}$. Using \equ{accretion_turb} as a proxy for the total turbulence in the clump, 
the ratio of this timescale to the inflow time can be evaluated: 
\be 
\label{eq:turb_diss}
\frac{t_{\rm diss}}{t_{\rm inflow}} \simeq 0.94 ~ \xv^{-1/3} \L1^{2/3} \Machv \epsilon_1^{-1/3}.
\ee
{\no}For a typical stream, the initial turbulence will dissipate in roughly the time it takes the stream 
to reach the central galaxy. At $\xv=0.3$ the ratio increases to $1.4$ for $\L1=1$, but if $\L1=0.5$ it 
is $\sim 0.9$.

\subsection{Formation of MP GCs}
\label{sec:GCs}

\begin{figure}
\begin{center}
\includegraphics[width =0.49 \textwidth]{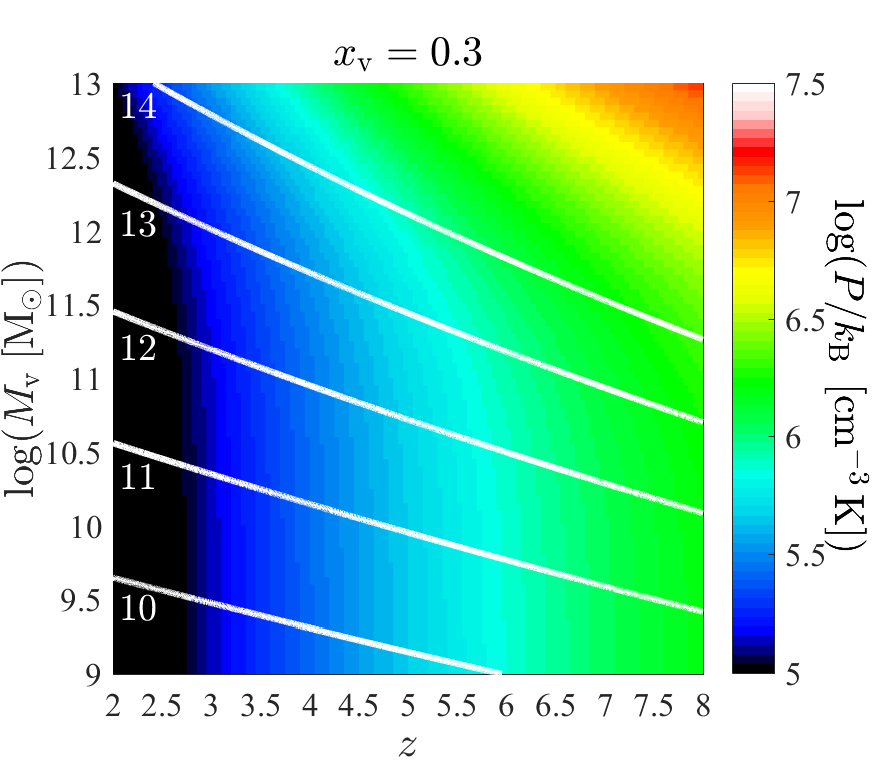}
\end{center}
\caption{Average pressure, $P=\rho c_{\rm eff}^2$, in the streams as a function of halo mass 
and redshift for parameters $\fst=\Machv=\T4=1$, $\L1=0.5$, and $\xv=0.3$. This corresponds to 
our fiducial values for the accretion rate, velocity and temperature of the stream, but a 
relatively narrow stream in the inner halo at $0.3\Rv$. The mean density is given 
by \equ{Rhos}, the sound speed by \equ{cs}, and the turbulent Mach number by \equ{accretion_mach} 
with fiducial efficiency, $\epsilon_{1}=1$, summed in quadrature with additional Mach 1 turbulence 
due to instabilities and feedback. Solid white lines mark the average mass histories of halos with 
present day masses ${\rm log}(\Mv(z=0)/\msun)=10,\,11,\,12,\,13$, and $14$, as marked. At $z\gsim 6$, 
the pressure is $P/k_{\rm B}\gsim 10^6\cmc \K$, as required to form GCs. At lower redshifts the pressure 
falls below this threshold. Note that this represents the mean pressure in the stream prior to the 
onset of gravitational instability, while the pressure in the collapsed clouds is expected to be larger.}
\label{fig:pressure} 
\end{figure}

\subsubsection{Gas Pressure in Streams}
We have established that streams at high redshift are unstable to fragmentation and that the cooling time 
is short enough to allow star-formation in the collapsing gas clouds. In order to form GCs in the collapsing 
clouds, they must reach high enough densities and pressures. GCs are expected to form only in very high pressure 
regions, with $P/k_{\rm B} > 10^6\cmc \K$ \citep{Elmegreen97,Kruijssen15}. The combined thermal plus turbulent 
pressure in the streams \textit{prior to gravitational collapse} is given by $P\simeq \rhos c_{\rm eff}^2$ (see 
\se{model2}). This is shown as a function of halo mass and redshift in \fig{pressure}. We assumed fiducial values 
for the stream velocity and temperature, the fraction of total accretion in the stream, and the efficiency of 
converting accretion energy into turbulent energy. In other words, $\Machv=\T4=\fst=\epsilon_1=1$. However, we 
assumed a relatively narrow stream, with $\L1=0.5$. The turbulent Mach number is given by $\mathcal{M}_{\rm turb,acc}$ 
from \equ{accretion_mach}, summed in quadrature with an additional $\mathcal{M}\sim 1$ to account for instabilities 
and feedback (see \se{turbulence}). The white lines in \fig{pressure} represent mass evolution tracks of halos with 
$z=0$ masses ${\rm log}(\Mv/\msun)=10,\,11,\,12,\,13$, and $14$. 

\smallskip
For the chosen parameters, the pressure in the streams exceeds $10^6\cmc \K$ at $z\gsim 6$ for all halo masses, and 
at $z\gsim 5$ for the progenitors of the most massive halos, $\Mv(z=0)\lsim 10^{14}$. Recall that the pressure plotted 
in \fig{pressure}, which represents the pressure in the streams prior to gravitational collapse, is not enough to maintain 
the streams in hydrostatic equilibrium when $\Mls>M_{\rm L,crit}$ (\equnp{Mlcrit}). The final pressure in the collapsed 
clouds prior to the onset of star-formation is thus likely to be even higher. We conclude that at $z\gsim 6$ the pressure 
in the streams in the inner halo is large enough to enable formation of GCs. However, this may not 
be true outside of $\sim 0.3\Rv$ due to the strong dependence of the density and pressure on position within the halo, 
$P \propto \rho \propto \xv^{-2}$. 

\subsubsection{Cluster Formation Efficiency}
\smallskip
In addition to large pressures, the formation of a GC requires very high densities. A typical GC with a mass of 
$2\times10^5$ and a half-mass radius of $3\pc$ has a mean density within this radius of 
$\sim 880\msun\pc^{-3}\sim2.6\times 10^4\cmc$. At $z=6$, at $0.3\Rv$, even a relatively narrow (and thus 
dense) stream with $\L1=0.5$ has a typical density of $\sim 13\cmc$ according to \equ{Rhos}, roughly a factor 
of 2000 below the densities in GCs. However, we cannot compare the mean density in the stream or in the pre-collapse 
cloud to the final GC density. Star-formation in a turbulent medium is a hierarchical process where the densest 
objects, namely bound stellar clusters and GCs, form at the highest density peaks within the cloud 
\citep[e.g.][]{Kruijssen12c,Hopkins13}. Furthermore, 
there is mounting evidence in the local Universe that the densest stellar clusters are significantly denser than 
the densest gas clouds \citep{Longmore14,Walker15}, suggesting that massive clusters are formed via hierarchical 
merging of smaller stellar clusters embedded in the parent gas cloud. 

\begin{figure*}
\begin{center}
\includegraphics[trim={0 0.5cm 2.7cm 0}, clip, width =0.3208 \textwidth]{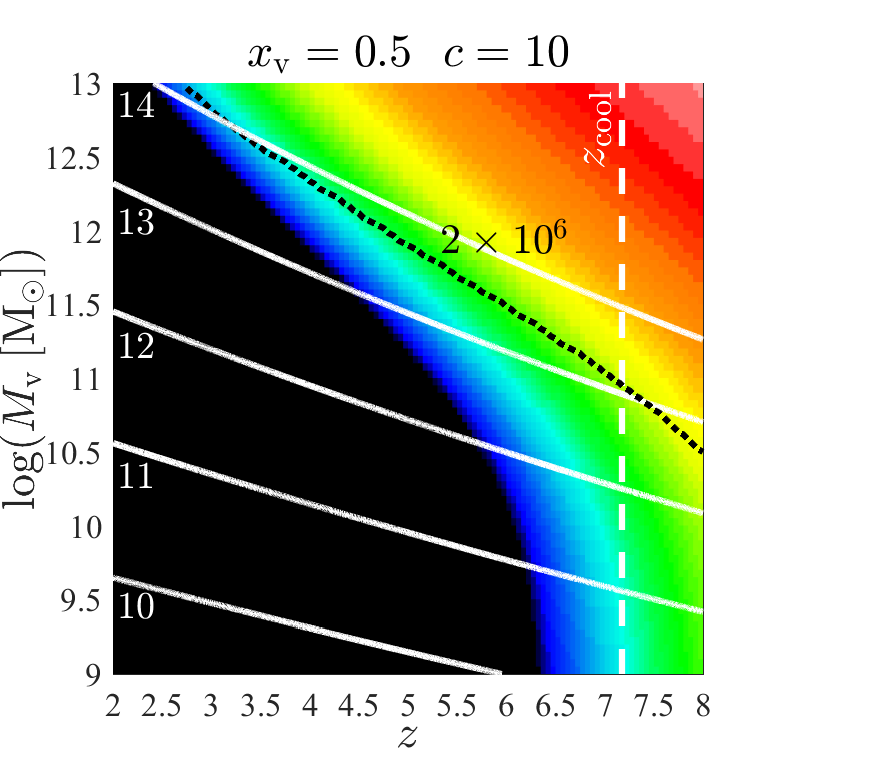}
\hspace{-0.15cm}
\includegraphics[trim={0 0.5cm 4.3cm 0}, clip, width =0.2783 \textwidth]{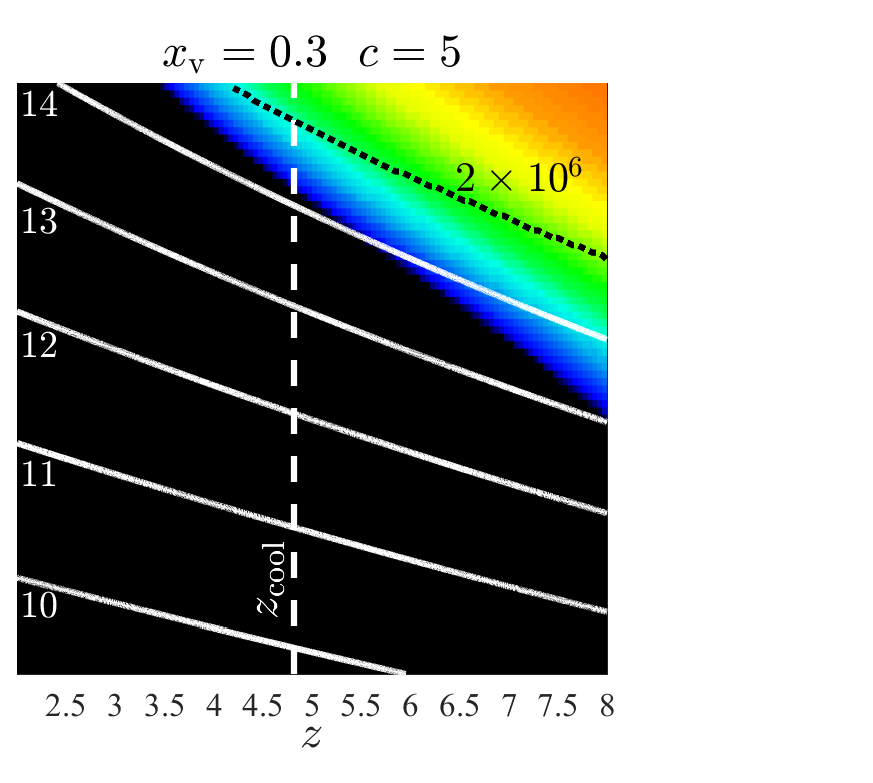}
\hspace{-0.15cm}
\includegraphics[trim={0 0.5cm 0.25cm 0}, clip, width =0.3854 \textwidth]{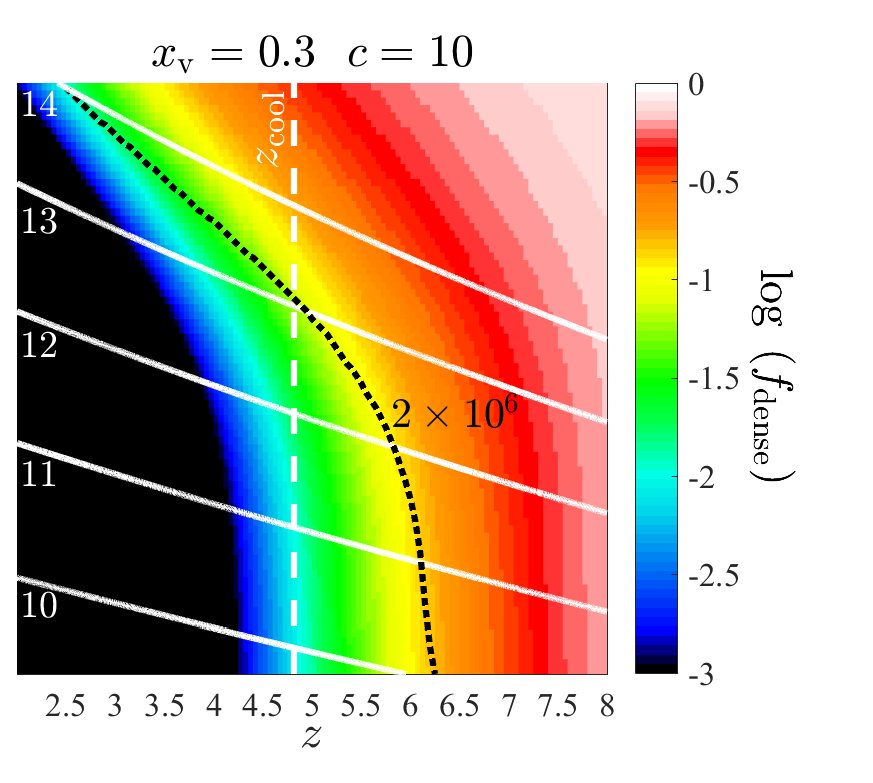}
\end{center}
\caption{Mass fraction within collapsed clouds at densities $\rho>\rho_{\rm GC}\sim 880\msun\pc^{-3}$, 
high enough to form GCs. As in \fig{pressure}, we assume stream parameters $\Machv=\fst=\T4=\epsilon_{1}=1$, 
and $\L1=0.5$. The turbulence in the cloud is given by \equ{accretion_mach} summed in quadrature with 
additional Mach 1 turbulence due to instabilities and feedback. The initial mean cloud density is given 
by \equ{Rhos}, but as the cloud contracts radially by a factor $c=5$ (center panel) or $10$ (left and 
right hand panels) its mean density increases by a factor $c^3$. 
The turbulence induces a density distribution given by \equs{dens_pdf} and \equss{sig_rho} with 
$b=0.5$. For the radial position in the halo, we show results for $\xv=0.5$ (left hand panel) and $\xv=0.3$ 
(center and right hand panels). Solid white lines mark the average mass histories of halos with present day 
masses ${\rm log}(\Mv(z=0)/\msun)=10,\,11,\,12,\,13$, and $14$, as marked. The vertical dashed lines mark 
$z_{\rm cool}$ from \equ{zcool}, the redshift above which the initial cooling time in the cloud prior 
to collapse is shorter than the inflow time of the stream to the central galaxy. For this equation we 
have assumed a clumping factor $\mathcal{C}_{10}=0.5$ and a metalicity $Z_2=2$ ($2\%$ solar). The black 
dotted line in each panel marks the contour where the total mass of gas with $\rho>\rho_{\rm GC}$ is 
$M_{\rm dense}=2\times 10^6\msun$, assuming the total gas mass in the collapsed cloud is given by the 
\textit{turbulent} Jeans mass, i.e. \equ{Mclump} multiplied by a factor $(1+\mathcal{M}^2_{\rm turb}/3)^{3/2}$. 
Above this curve the clouds have enough dense gas to form GCs. Due to the relatively 
low stream densities in the outer halo, $\xv\gsim 0.5$, the cooling time is long and we have $z_{\rm cool}\gsim 7$. 
Furthermore, even with very large contraction factors of $c=10$ and a narrow stream, there is not enough dense 
gas to form GCs except at very high redshifts $z>7$. With contraction factors of $c=5$, the same is true even 
near the inner halo, at $\xv=0.3$. However, at $\xv=0.3$ with $c=10$, roughly $10\%$ of the cloud mass is dense 
enough to produce GCs in halos with virial masses down to $10^{9}\msun$ at $z>5.5$, while $z_{\rm cool}\sim 4.8$. 
Such large contraction factors likely require loss of angular momentum in the collapsing cloud, which can occur 
in the inner halo when counter-rotating streams collide (see text). 
}
\label{fig:dense_mass} 
\end{figure*}

\smallskip
Within this framework, one can evaluate the fraction of star-formation occuring in bound clusters, 
referred to as the cluster formation efficiency, or CFE \citep[e.g.][]{Bastian08,Goddard10}, 
based on the density and pressure within the parent cloud\footnote{There is still some debate 
in the literature regarding the origin of the CFE, with some claiming that it does not depend on 
the properties of the host galaxy or of the parent cloud \citep{Fall12,Chandar15,Chandar17,Mulia16}. 
However, we here adopt the theoretical framework of \citet{Kruijssen12b} whereby the CFE depends on 
the density and pressure in the parent cloud.} \citep{Kruijssen12b,Adamo15,Johnson16}. For a 
$10^{10}\msun$ halo at $z=6$ with the same parameters as in \fig{pressure}, the streams 
have pressure $P/k_{\rm B} \sim 10^6\cmc \K$, density $\rho \sim 0.44 \msun \pc^{-3}$, and column density 
$\Sigma\sim 100\msun \pc^{-2}$. These properties are very similar to those of the Fornax model of \citet{Kruijssen15} 
(Table 1), which had $P/k_{\rm B} = 1.06\times 10^6\cmc \K$, $\rho=0.82 \msun \pc^{-3}$, $\Sigma=103 \msun \pc^{-2}$, 
and a CFE of $\Gamma \sim 0.38$. In our model, the density and pressure in the collapsed clouds prior to the 
onset of star-formation are likely to be much higher than the typical stream values, so CFE values of $\sim 0.4$ 
or larger are entirely plausible. 

\smallskip
The CFE can be used to determine the maximal cluster mass that forms within the parent cloud, 
$M_{\rm cl,max}\simeq \Gamma \epsilon_{\rm SF} \Mj$ where $\epsilon_{\rm SF}$ is the fraction of 
the cloud mass that turns into stars \citep{RC17}. For $\Gamma\sim 0.3$, $\epsilon_{\rm SF}\sim 0.1$, 
and $\Mj\sim 2\times 10^7\msun$ (\equnp{Mclump}), we infer a maximal cluster mass of order $6\times 10^5\msun$ which is 
again consistent with the maximal cluster mass in the Fornax model of \citet{Kruijssen15}. For halos with $\Mv>10^{10.5}\msun$ 
at $z\sim 6$, which are the progenitors of $>10^{12}\msun$ halos at $z=0$, we expect larger CFE values due to stronger 
gravitational instability as well as larger Jeans masses due to stronger turbulence (see \se{model2}). Both of these will 
yield larger cluster masses.

\subsubsection{Dense Gas Fraction}
\smallskip
Notwithstanding the above discussion, we defer a more detailed evaluation of the CFE and the maximal cluster 
mass in streams to future work, focussing here instead on the simpler question of under what conditions a 
collapsed cloud will have enough gas at high enough densities to lead directly to the formation of a GC. Based 
on the above discussion, this is a sufficient but not necessary condition for GC formation. As previously 
stated, a typical GC is $\sim 2000$ times denser than a typical stream. For the mean density in the 
collapsed cloud to reach these values, a radial contraction factor of $c=\Lj/R_{\rm cl}>13$ is required, 
where $R_{\rm cl}$ is the radius of the collapsed cloud. At lower redshifts and at larger halocentric distances, 
the discrepancy is larger. While we have argued in the previous section that the relatively long cooling times 
in the clouds compared to their free-fall timescale can lead to significant contraction factors before the onset 
of star-formation, a contraction factor of $c>10$ may still be difficult to achieve due to angular momentum support. 

\smallskip
However, as described above it is not necessary for the mean density in the cloud to be as dense as a GC, but only 
that the highest density peaks are dense enough. The volume-weighted PDF of the mass overdensity in an isothermal 
turbulent medium is well described by a lognormal distribution \citep[e.g.][]{Federrath10,Konstandin12} 
\be 
\label{eq:dens_pdf}
d{\rm P}/ds = \frac{1}{\sqrt{2\pi \sigma_{\rm \rho}^2}s}{\rm exp}\left(\frac{\left({\rm ln}(s)+0.5\sigma_{\rm \rho}^2\right)^2}{2\sigma_{\rm \rho}^2}\right),
\ee
{\no}where $s=\rho/{\bar {\rho}}$ is the overdensity, and 
\be 
\label{eq:sig_rho}
\sigma_{\rho}^2 = {\rm ln}(1+b^2\mathcal{M}^2),
\ee
{\no}where $\mathcal{M}$ is the turbulent Mach number, and $b$ depends on the ratio of compressive 
to solenoidal modes in the turbulence. For a ``natural" mixture of modes $b\sim 0.5$, while it approaches 
$1$ for more compressive forcing \citep{Federrath10}. We can use this expression to estimate the fraction 
of mass that will be dense enough to form a GC, i.e. $\rho\gsim 880\msun\pc^{-3}$. 


\smallskip
We assume that the cloud begins with mean density $\rhos$ (\equnp{Rhos}) and then contracts radially 
by a factor of $c$, so that its final density is ${\bar {\rho}}\sim c^3 \rhos$. The turbulence in the 
initial cloud is given by $\mathcal{M}_{\rm turb,acc}$ from \equ{accretion_mach}, summed in quadrature with 
an additional $\mathcal{M}\sim 1$ to account for instabilities and feedback, as in \fig{pressure}. As 
the cloud contracts, its internal turbulence may be amplified. In the absense of any dissipation or cooling, 
$\sigma_{\rm turb}^2 \propto GM/R$, so the turbulent Mach number scales as $\mathcal{M}_{\rm turb} \propto c^{1/2}$. 
However, in practice this maximal enhancement is rarely seen in simulations since turbulence dissipates 
more rapidly as the cloud contracts. As a conservative estimate, we ignore the possible enhancement of 
turbulence and assume that it remains constant during the contraction. This may result in an underestimate 
of the dense gas fraction. 

\smallskip
In \fig{dense_mass}, we show the fraction of gas in the collapsed cloud which is denser than $\rho_{\rm GC}=880\msun\pc^{-3}$ 
as a function of halo mass and redshift. The different panels explore different values of the contraction factor, 
$c=5$ and $10$, and the halocentric distances, $\xv=0.3$ and $0.5$. All other parameters are the same as in \fig{pressure}, 
namely $\Machv=\T4=\fst=\epsilon_1=1$ and $\L1=0.5$. The vertical dashed line shows $z_{\rm cool}$ from \equ{zcool}, 
assuming a clumping factor of $C_{10}=0.5$ and a metalicity of $Z_2=2$. The white solid lines in each panel represent 
halo mass evolution tracks, as in \fig{pressure}. 

\smallskip
GCs are expected to undergo mass loss of a factor of $\sim 2-10$ from their formation until 
$z=0$ \citep{Kruijssen15,MBK17}. The formation mass of a typical GC should thus be in the 
range $0.5-2\times 10^6\msun$. In practice, since our model predicts that GCs form directly 
in the halo rather than in the central disc, the actual mass loss may be on the low side 
\citep{Kruijssen15}, so a typical GC with mass $2\times 10^5\msun$ at $z=0$ may have 
had a formation mass as low as $\sim 5\times 10^5\msun$. Nevertheless, we hereafter adopt 
the conservative estimate that the formation mass of a GC is $\sim 2\times 10^6\msun$, and 
require at least this much gas to be above $\rho_{\rm GC}$. This can also allow for the formation 
of multiple clusters with a range of masses in a given parent cloud.

\smallskip 
We can estimate the total mass of gas with $\rho>\rho_{\rm GC}$ by multiplying the fraction of dense 
gas shown in \fig{dense_mass} by the \textit{turbulent} Jeans mass (see \se{model2}). The black dotted 
line in each panel of \fig{dense_mass} marks the contour of $2\times 10^6\msun$ of gas with 
$\rho>\rho_{\rm GC}$ in a collapsed cloud. Above this line, there is enough dense gas to form a GC. 
We see that in all cases, at $z>z_{\rm cool}$ this contour roughly corresponds to the contour of 
$f_{\rm dense}\sim 0.1$. 

\smallskip
From the left-hand panel of \fig{dense_mass} we learn that in the outer halo, at $\xv>0.5$, even with very large 
contraction factors of $c=10$ only the progenitors of $\sim 10^{14}\msun$ halos at $z=0$ have enough dense gas to 
form GCs at $z\sim 6$. Even at $z\sim 8$, only progenitors of $\gsim 10^{13}\msun$ halos have enough dense gas to 
form GCs. We conclude that GC formation in streams is unlikely to occur in the outer halo, which was the same conclusion 
we reached when considering the pressure in the streams in \fig{pressure}. 

\smallskip
From the center panel of \fig{dense_mass} we see that even in the inner halo, at $\xv=0.3$, a contraction factor 
of $c=5$ will not result in direct collapse of a GC within the collapsed cloud, except in extremely massive 
halos with $\Mv>10^{12.5}\msun$ at $z\gsim 6$. 

\smallskip
In the right-hand panel of \fig{dense_mass}, we examine the case of $\xv=0.3$ and $c=10$. In this case, 
$z_{\rm cool}\sim 4.8$. At $z\gsim 5.5$, for all halos with $\Mv\gsim 10^9\msun\sim M_{\rm \nu=2}$, 
$\sim 10\%$ of the mass of the collapsed cloud has densities $\rho>\rho_{\rm GC}$. Furthermore, there 
is enough dense gas in the collapsed cloud to directly form a GC with $\sim 2\times 10^6\msun$ at $z\sim 6$. 
We conclude that if the collapsing clouds can reach contraction factors of $c\sim 10$, then MP GC formation 
will occur in cold streams in the inner $\sim 30\%$ of the halo, outside the central galaxy. 

\subsubsection{Stream Collisions}
\smallskip
What now remains is to ask whether such a large contraction factor is possible. As noted above, since the 
gas stream is at least partially supported by rotation, the cloud contraction will be limited by angular 
momentum support at some point. While the degree of angular mometum support, and thus the maximal contraction 
factor, are unknown, it seems unlikely that the clumps will be able to contract in size by a factor $c=10$ without 
losing some angular momentum. Fortunately, there is a natural mechanism for this in the inner halo. As described 
in \se{fil_obs}, interior to $\sim 0.3\Rv$ the streams interact with each other and with the central disc as they 
form an extended rotating ring and spiral towards the central galaxy within an orbital time \citep{Danovich15}. 
While most of the streams are co-rotating, thus efficiently transporting angular momentum to the growing disc, 
simulations indicate that $\sim 30\%$ of the mass flowing into this interaction region is actually counter-rotating 
\citep{Danovich15}. A counter-rotating stream can collide with another stream in this region. Such a collision 
could significantly reduce the angular momentum of the stream and of the collapsing clouds, as required. Furthermore, 
the collision itself could significantly increase the densities of the collapsing clouds. The relative velocity 
of the collision is of the order of the stream velocity, $\Machv\Vv$, and thus has a Mach number of 
\be 
\label{eq:collision}
\mathcal{M}_{\rm coll} \simeq 8~\M10^{1/3} \z7^{0.5} \T4^{-0.5}.
\ee
{\no}The very short cooling times in the streams following such a collision should result in an isothermal shock 
(see \citealp{Cornuault16} for a discussion on strong isothermal shocks in cold streams as they enter the virial 
radius of a hot halo), resulting in an increase in the stream density by a factor $\sim \mathcal{M}_{\rm coll}^2$. 
For a $10^{12}\msun$ halo at $z=6$ this is comparable to the density increase in a cloud with a radial contraction 
factor of $c\sim 10$, though for lower mass halos and lower redshift the density increase is smaller. 
The mean density in the collapsing clouds may increase by a similar factor as well, allowing GC formation at lower 
global contraction factors\footnote{We note that this scenario bears certain qualitative similarities to 
the scenario proposed by \citet{Fall85}, except that here the dense clouds are confined by the ram pressure of stream 
collisions rather than the thermal pressure of a hydrostatic halo at the virial temperature, though both these pressures 
have comparable magnitudes.}. We conclude that such stream collisions in the inner halo may act as the trigger for MP 
GC formation. 

\smallskip
\section{Summary and Predictions of the Model}
\label{sec:model_disc}

\begin{table*}
\centering
\begin{tabular}{@{}lccccc}
\multicolumn{5}{c}{Model Parameters} \\
\hline
Parameter & Meaning & Reference eqn. & Fiducial Value & Plausible Range \\
\hline
$\fs$         & Fraction of average halo accretion along a single stream     & \equ{Mdot}    & $0.33$    & $0.1-1.0$ \\ 
$\Machv$      & Ratio of stream velocity to halo virial velocity             & \equ{Vdef}    & $1.0$     & $0.5-2.0$ \\
$\Ls$         & Ratio of stream radius to dark matter filament virial radius & \equ{spin}    & $0.1$     & $0.02-0.20$ \\
$\Ts$         & Stream gas temperature                                       & \equ{cs}      & $10^4 \K$ & $0.8-3\times 10^4\K$  \\
$\epsilon$    & Turbulence driving efficiency                                & \equ{heitsch} & $0.01$    & $0.01-0.10$ \\ 
$\mathcal{C}$ & Clumping factor in the streams                               & \equ{cool}    & $5$       & $5-20$ \\ 
$Z_{\rm s}$   & Stream gas metalicity                                        & \equ{cool}    & $0.02$    & $0.003-0.03$ \\
\hline
\end{tabular}
\caption{}
\label{tab:params}
\end{table*}

\begin{figure}
\begin{center}
\includegraphics[trim={2.0cm 1.5cm 0 0}, clip, width =0.49 \textwidth]{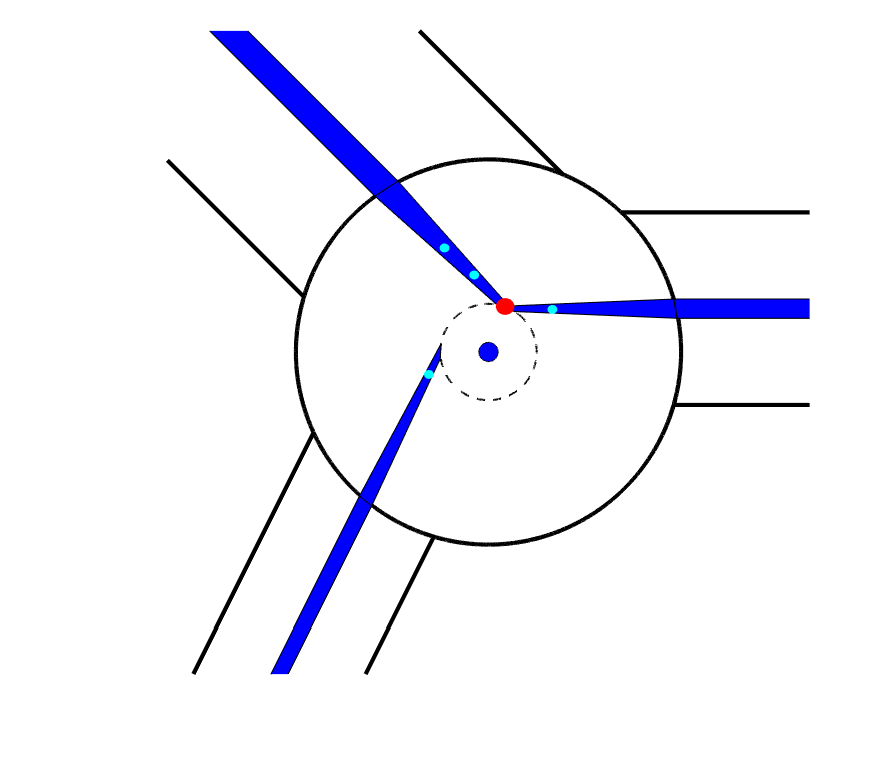}
\end{center}
\caption{Schematic diagram summarizing the main components of the model. A massive dark matter halo, 
containing a star-forming galaxy within the inner $0.05-0.1\Rv$, is fed by 3 cylindrical dark matter 
filaments with radii $r_{\rm v,Fil}\sim 0.4-0.8\Rv$. Each dark-matter filament contains a narrow 
stream at its center, whose radius is roughly $\Rs\sim 0.05-0.1r_{\rm v,Fil}$. Within the virial radius, 
the dark matter filaments mix in with the halo and virialize, while the gas streams form a conical 
shape and plunge towards the halo center. These gas streams have impact parameters of the order 
$\sim 0.3\Rv$, marked by a dashed circle. Within this region the streams interact with each other 
and with the central disc galaxy until they eventually join the disc. Some streams may counter-rotate 
with respect to the total angular momentum, resulting in a stream collision within the interaction 
region. This will lead to the formation of a massive, bound GC. Less massive clusters and field stars 
can form within the streams themselves farther out in the halo.}
\label{fig:cartoon} 
\end{figure}

\subsection{Stream Fragmentation}
\smallskip
\Fig{cartoon} presents a schematic diagram highlighting the main ingredients of our model. 
Massive halos at high redshift are fed by cosmic web filaments of dark matter which contain 
dense streams of cold, $\sim 10^4\K$, gas in their centers. Outside the halo the filaments 
and streams are roughly isothermal cylinders. Inside the halo the dark matter filaments mix 
in with the surrounding medium and virialize, while the streams remain cold and coherent as 
they penetrate towards the inner $\sim 0.3\Rv$ in a conical shape. This element of the model 
is not new, and has been discussed in several previous works \citep[e.g.][]{Danovich15}. However, 
our model discusses in detail for the first time gravitational instability and fragmentation of 
the streams. Along the way, the streams become gravitationally unstable to both large-scale 
cylindrical instabilities and local Jeans instabilities. They fragment into bound clumps with 
masses $M_{\rm c}\sim 2-5\times 10^7~(10^8)\msun$ in halos with $\Mv \lsim 10^{11}~(\sim 10^{12})\msun$. 
At high redshift, typically $z>z_{\rm cool}\sim 4$ though possibly at lower redshifts if the 
metalicity is more than $1-2\%$ solar, these clumps can 
cool and form stars before reaching the central galaxy. This is seen in cosmological simulations 
(\figs{cosmo} and \figss{profiles}), and may have been observed in a few instances (see \se{fil_2}). 
Little or no dark matter is expected to be bound to these clumps despite their location at distances 
of up to tens of $\kpc$ from the central galaxy. While many of these clumps may eventually be destroyed 
by the tidal forces of the central galaxy, they can contribute to the growth of the metal-poor population 
in the stellar halo ($Z\lsim 0.03Z_{\odot}$, \fig{metals}) and may be observed as Ly$\alpha$ emitters. 

\smallskip
There are seven parameters that define the various aspects of our model. These are summarized in \tab{params}, 
along with the equation where they first appear and with their fiducial values and plausible ranges, based on the 
arguments laid forth in \se{model}. The gravitational stability of cold streams depends on $\fs$, $\Machv$, and $\Ts$, 
with faster, hotter and less prominent streams being more stable (\equsnp{LJeans} and \equssnp{Mlcrit}). Pushing these 
to their extreme values, our model predicts that the minimal halo mass where streams are likely to be gravitationally 
unstable at $z=6$ is in the range $\sim 2\times 10^9-2\times 10^{11}\msun$, while for typical values halos with 
$\Mv\gsim 2\times 10^{10}\msun$ at $z=6$ should host unstable streams. Whether this instability is able to manifest 
itself in the halo, before the streams reach the central galaxy, depends primarily on the product $\Machv \Ls$ (\equsnp{tff}, 
\equssnp{tcoll}, and \equssnp{zcool}). For typical values, clumps collapse and cool before reaching the galaxy, though 
if a stream is both very wide and rapidly inflowing this may not be the case. The thermal Jeans mass in the streams 
depends on stream width, $\Ls$, and on position within the halo (\equnp{Mclump}). For the full range of plausible 
$\Ls$ values and for halocentric distances of $(0.3-1)\Rv$, the thermal Jeans mass is $\sim 4\times 10^6-10^8\msun$, 
though it is expected to typically be a few times $10^7\msun$.

\smallskip
The level of turbulence in the streams depends on the first five parameters in \tab{params} (\equnp{accretion_mach}). 
The turbulent Mach number increases for slower, wider, colder, and more prominent streams, with a larger driving efficiency. 
In our fiducial model, turbulence played a significant role only in the most massive halos, $\Mv\sim 10^{12}\msun$ at $z=6$, 
while for less massive halos and at lower redshifts thermal pressure dominated over turbulent pressure. Therefore, if the 
parameters are such that the turbulence is even lower, this will not alter our main conclusions. On the other hand, if we 
take extreme values for all parameters to make the turbulence as large as possible ($\fs=1$, $\Machv=0.5$, $\Ls=0.2$, 
$\Ts=0.8\times 10^4\K$, $\epsilon=0.05$), the turbulent pressure can dominate even in $10^{10}\msun$ halos. This would 
yield larger density and pressures peaks in the streams, thereby strengthening the case for globular cluster formation. 
However, such a parameter combination seems unlikely, and we do not expect turbulence to be significant in halos less 
massive than a few times $10^{11}\msun$.

\smallskip
Finally, the redshift above which the streams can cool to $T\lsim 100\K$ and form stars before reaching the central 
galaxy depends on the product of $\mathcal{C} Z_{\rm s}$ (as well as $\Machv \Ls$ as described above, \equnp{zcool}). 
A stream with both very low metalicity and very small clumping factor will not be able to cool and form stars at $z\sim 5-6$ 
as in our model. However, our fiducial clumping factor of $\mathcal{C}\sim 5$ is at the low end of the expected range, 
and cosmological simulations suggest that streams feeding massive galaxies have been enriched to roughly $1-3\%$ solar 
metalicity by $z\sim 6$, so we expect that for typical cases streams should be able to cool and form stars at $z\gsim 5$. 
The plausible range for these parameters may even lead to cooling and star-formation at somewhat lower redshifts.

\subsection{Globular Cluster Formation}
\smallskip
At $z\gsim 6$, in the inner $\sim 0.3\Rv$ for a relatively narrow (though by no means extreme) stream with 
$\Ls\sim 0.05$, the pressure in the streams is large enough to enable GC formation (\fig{pressure}). The cluster formation 
efficiency (CFE) is expected to be $\gsim 0.3-0.4$, yielding maximal cluster masses of $\sim 6\times 10^{5}\msun$ in halos 
with $\Mv\lsim 10^{10}\msun$. The maximal cluster mass may increase by a factor of $\gsim 10$ in halos with $\Mv\sim 10^{12}\msun$, 
both because of an increased turbulent Jeans mass and an expected increase of the CFE. These values are consistent with 
previous estimates of GC masses at formation in models where GCs form in dense GMCs within high redshift discs \citep{Kruijssen15}. 
The low metalicity in the cold streams results in a low metalicity of $\lsim 0.03Z_{\odot}$ for these clusters, 
similar to MP GCs. However, due to the strong dependence of the density and pressure on stream width, it is 
unlikely that regular star formation in the streams will result in GC formation for $\Ls\gsim 0.1$.

\smallskip
The inner $\sim 0.3\Rv$ consists of a messy interaction region where the streams join the disc. 
In this region, collisions between counter-rotating streams can result in densities and pressures 
large enough to form GCs even if $\Ls\sim 0.1$ (\fig{dense_mass}), and may be the dominant formation 
channel of MP GCs in our model. These proto-GCs may have masses of $\lsim 10\%$ the mass of their parent 
cloud, namely a few times $10^6\msun$ ($10^7\msun$ for the most massive halos), and be essentially devoid 
of dark matter. Their masses then decrease by a factor of $3-10$ by $z=0$, yielding masses typical of observed 
GCs. 

\smallskip
Our model may allow for formation of GCs at relatively low redshifts, down to $z\sim 4.7$ for the progenitors 
of $10^{13}\msun$ halos at $z=0$, and down to $z\sim 4$ for the progenitors of $10^{14}\msun$ halos. The latter 
is assuming larger clumping factors due to larger turbulent velocities, in order to allow the collapsing clouds 
to cool before reaching the halo centers. This likely results in higher metalicities for the GCs formed in more 
massive halos, owing to the overall larger metalicities at lower redshifts. Additionally, those GCs that form at 
lower redshift form in more massive clouds (\equnp{Mclump}). A trend between the mass of individual MP GCs and 
the mass of their host halo is indeed observed \citep{Harris17}. This is also qualitatively consistent with the 
observed correlation between the mass and metalicity of MP GCs, the ``blue tilt'', since more massive halos at lower 
redshift will have higher metalicity gas. This feature is not observed for MR GCs \citep[e.g.][]{Brodie06}, possibly 
hinting at a different formation environment for the two populations. 

\smallskip
The kinematics of the MP GC population is a result of the kinematics of the streams in the inner 
$\sim 0.3\Rv$ where we estimate fragmentation is most likely to occur. As discussed in \se{fil_obs}, 
this is the region where the angular momentum is transferred from the streams to the disc due to strong 
torques, and the streams form an extended, high angular momentum, ring-like structure. Formation 
of MP GCs in these conditions may help explain the rapid rotation velocities and velocity anisotropies 
observed in the MP GC population in the outer halos \citep{Pota15a}. 

\smallskip
Our picture of MP GC formation in the inner halo allows us to place constraints on two important observables: 
the ratio of the mass of a MP GC system (GCS) to that of its host halo at $z=0$, and the ratio of the extent 
of a GCS to the virial radius of its host halo at $z=0$. Both of these are observed to be constant over several 
orders of magnitude in halo mass \citep{Harris17,Forbes17}. 

\begin{figure}
\begin{center}
\includegraphics[width =0.49 \textwidth]{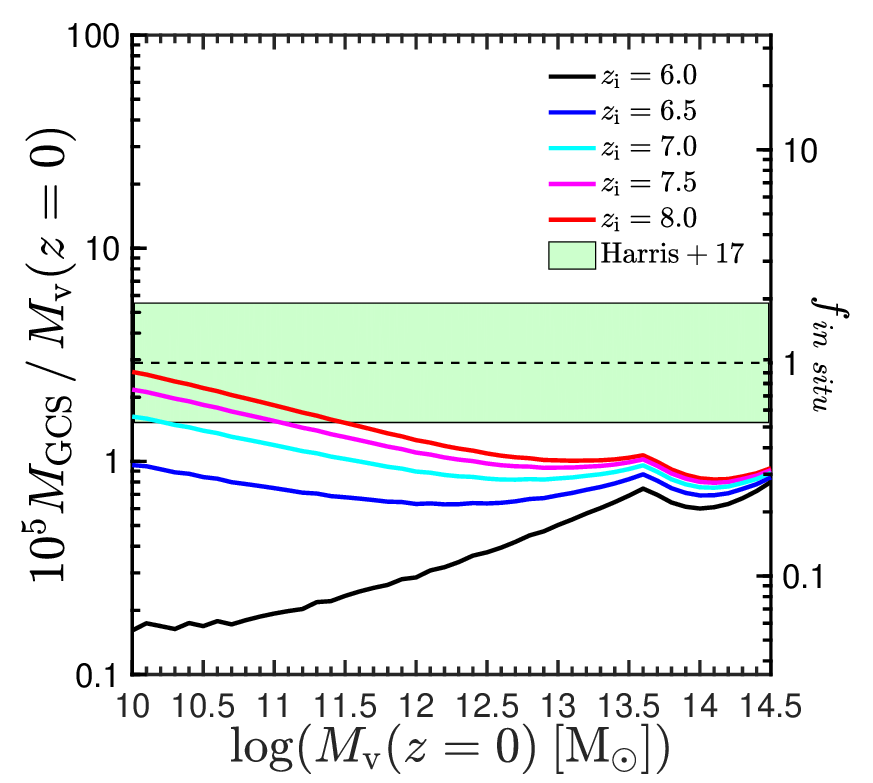}
\end{center}
\caption{Estimate of the ratio of the total mass in MP GCs that formed within the halo at $z\sim 6$ to the 
mass of the host halo at $z=0$, as a function of host halo mass, according to our model. The solid curves 
represent our model, which assumes that all counter-rotating gas that enters $0.3\Rv$ between an initial 
redshift $z_{\rm i}$ a final redshift $z_{\rm 10\%}$ forms MP GCs at $10\%$ efficiency. $z_{\rm 10\%}$ is 
defined as the redshift where $f_{\rm dense}$ from \fig{dense_mass} falls below $10\%$, and we have assumed 
the same model parameters as in the right-hand panel of \fig{dense_mass}. In halos with $\Mv<10^{12}\msun$ 
all gas entering the virial radius is assumed to penetrate down to $0.3\Rv$, while in more massive halos 
the penetration is only $\sim 50\%$ \citep{Dekel13}. The counter-rotating gas represents $30\%$ of the total 
gas entering $0.3\Rv$, based on simulations \citep{Danovich15}. The MP GCs then decrease in mass by a factor 
of 10 by $z=0$. Different colour lines show results for different values of $z_{\rm i}$, ranging from 6 to 8. 
The dashed line and shaded region show the mean ratio and $1-\sigma$ scatter from \citet{Harris17} for the total 
mass of MP GCs to the host halo mass, $M_{\rm GCS}/\Mv \sim 2.9\times 10^{-5} \pm 0.28~{\rm dex}$. At low halo 
masses, where accretion of \textit{ex situ} GCs via mergers is expected to be negligible, our model is in 
good agreement with the observations. At higher halo masses we predict a larger fraction of accreted GCs. For 
MW mass halos, $\Mv\sim 10^{12}\msun$, we predict that $\sim 30\%$ of the MP GCs were formed in halos at $z\gsim 6$, 
in good agreement with the prediction of \citet{MBK17}. The kink in the model curves at $\Mv\sim 10^{13.5}\msun$ 
marks halos that become more massive than $10^{12}\msun$ at $z>z_{\rm 10\%}$.}
\label{fig:MGCS} 
\end{figure}

\smallskip
An upper limit to the total mass of the MP GCs that formed \textit{in situ} in the host halo as a result 
of collisions between counter-rotating streams, rather than having been accreted at lower redshift, can 
be obtained as follows. As stated above, simulations indicate that $\sim 30\%$ of the gas mass entering 
$0.3\Rv$ is counter-rotating. We assume that all the counter-rotating gas that enters $0.3\Rv$ between 
an initial redshift $z_{\rm i}$ and the redshift where $f_{\rm dense}$ from \fig{dense_mass} falls below 
$10\%$, $z_{\rm 10\%}$, contributes to formation of MP GCs at $10\%$ efficiency. At these high redshifts, 
the halo mass as a function of redshift is well approximated by $M_{\rm v}(z)\simeq M_{\rm v}(z_0){\rm exp}(-0.79[z-z_0])$ 
\citep{DM14}. For all halos of interest, $z_{\rm 10\%}\sim 5-6$. For a given $z_{\rm i}$, the fraction of 
mass accreted onto a halo in the redshift range $z_{\rm i}-z_{\rm 10\%}$ is independent of halo mass. For 
$z_{\rm i}\sim 7-8$, this fraction is $\sim 0.55-0.90$. Thus the efficiency of GC formation relative to the 
total baryonic mass in the halo at $z_{\rm 10\%}$ is $\sim 0.1\times 0.3\times (0.55-0.90) = 0.0165-0.027$, 
independent of halo mass. In his empirical model of hierarchical merging of globular cluster systems, \citet{MBK17} 
found that formation efficiencies of $\sim 1.4-2.8\%$ for MP GCs in halos at $z=6$, combined with GC mass 
loss by a factor of $\sim 10$ until $z=0$, can account for the present day masses of GC systems. 

\smallskip
If we assume additional mass loss by a factor of $\sim 10$ until $z=0$, we have $\sim 1\%$ of the counter-rotating 
mass entering $0.3\Rv$ from $z_{\rm i}$ to $z_{\rm 10\%}$ winds up in MP GCs at $z=0$. Note that this represents 
mass loss from the entire GCS rather than from an individual GC, and thus includes GC disruption as well. This is 
$0.01\times 0.3\times p=0.003p$ of the total gas mass or $0.003p\times \fb \sim 5.1\times 10^{-4}p$ of the total 
mass accreted onto the halo in this interval, where $p$ is the fraction of gas crossing the virial radius that 
penetrates down to $0.3\Rv$. In halos with virial mass $\Mv\lsim 10^{12}$ we expect $p\sim 1$, while in more 
massive halos we expect $p\sim 0.5$ due to the presence of a hot halo out to the virial radius \citep{Dekel13,DM14}. 
This is only relevant for the progenitors of very massive halos with $\Mv>10^{13.5}\msun$ at $z=0$, since lower mass 
halos are always less massive than $10^{12}\msun$ at $z>z_{\rm 10\%}$ (\fig{dense_mass}). 

\smallskip 
In \fig{MGCS} we show the results of this simple model for different values of $z_{\rm i}=6-8$, for the same model 
parameters as in the right-hand panel of \fig{dense_mass}. For each chosen $z_{\rm i}$ we show the ratio of the total 
GCS mass to the host halo mass at $z=0$ as a function of host halo mass. The kink at $\Mv\sim 10^{13.5}$ in all model 
curves is due to the progenitors of these massive halos crossing the $10^{12}\msun$ threshold at $z>z_{\rm 10\%}$. For 
comparison we show the recent result of \citet{Harris17}, who found a ratio of $\sim 2.9\times 10^{-5} \pm 0.28~{\rm dex}$. 
For $z_{\rm i}\gsim 7.5$, our results for low mass halos with $\Mv<10^{11}\msun$ are in very good agreement with the 
observations, suggesting that a large fraction of their MP GCs could have formed via this mechanism. At higher masses 
we underproduce the total mass of the MP GCS. But this is to be expected, since we are only accounting here for MP GCs 
that formed as a result of stream collisions in the halo. The remaining GCs are \textit{ex situ} in origin, accreted onto 
the halo during mergers of sub-halos\footnote{As discussed, we also expect some MP GC formation in streams feeding massive 
halos even in the absense of collisions. The contribution of these clusters to the total population depends on both the CFE 
and the total amound of star formation in the streams, and will be the subject of future work.}. 
We find that the fraction of mass in \textit{ex situ} MP GCs accreted during mergers increases with 
halo mass, because more massive halos accumulated a larger fraction of their total mass through mergers of halos with 
$\Mv>10^9\msun$ at $z<6$, the threshold for \textit{in situ} MP GC formation in our model. For a MW mass galaxy, with 
$\Mv(z=0)\sim 10^{12}\msun$, we find that $\sim 30\%$ of the total GC mass was formed in the halo of the main progenitor 
at $z\gsim 6$. This is in good agreement with the estimates of \citet{MBK17} who constrained this value using halo merger 
trees. In his model, he assumed that every halo above $\sim 10^9\msun$ at $z=6$ formed MP GCs in the halo with a total 
mass proportional to the halo mass at $z=6$, but did not propose a physical mechanism by which this could occur. Our model 
provides such a physical mechanism, and forms the basis for formation of MP GCs directly in the halos of massive galaxies 
at $z\sim 6$. Hierarchical models can then be used to assess the accretion of \textit{ex situ} GCs via mergers of sub halos.

\begin{figure}
\begin{center}
\includegraphics[width =0.49 \textwidth]{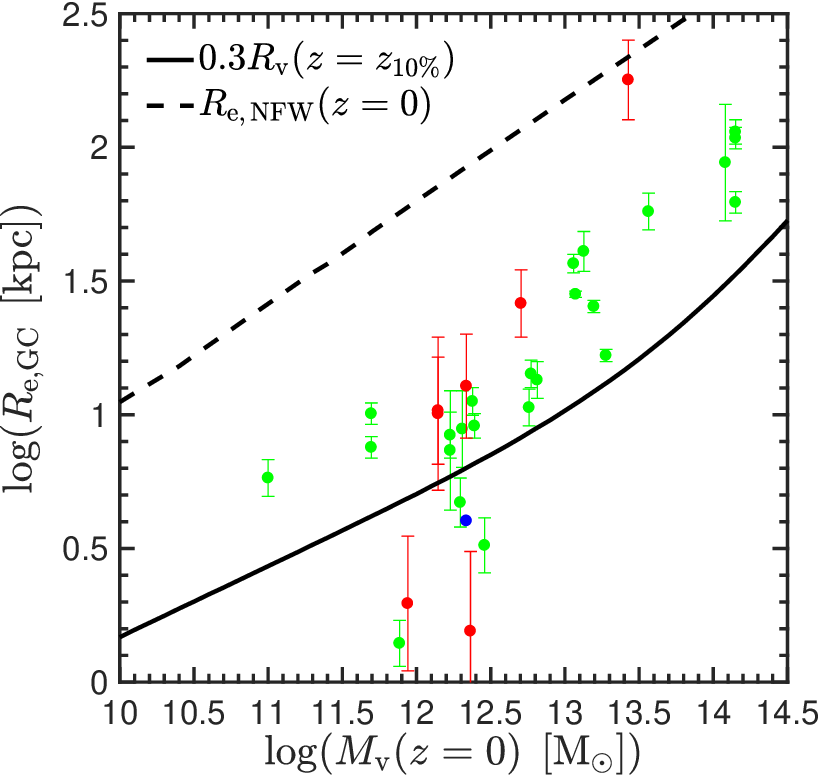}
\end{center}
\caption{Radial extent of MP GCs at $z=0$ as a function of host halo mass. The points show data from \citet{Forbes17}, 
specifically their figure 5. The red points are galaxies from the study of \citet{Hudson17}, the green points are other 
data from the existing literature including three UDGs, and the blue point is the MW. Mesurement data represent the 
half-number radius of the GCS. The solid black line shows the prediction from our model for the radial extent of those 
MP GCs that formed within the halos of their host galaxies at $z\sim 6$. We assume that the radial extent of these  
MP GCs is $\sim 30\%$ of the virial radius of their hos halo at $z_{\rm 10\%}$, defined as in \fig{MGCS}. Our model 
systematically lies $\sim 0.1-0.3~{\rm dex}$ below the observations, though the slope of our predicted relation is 
very similar to the observed slope, including the steepening above $10^{13}\msun$. The larger observed sizes are due 
to additional \textit{ex situ} GCs that were accreted onto the halo via mergers and lie at larger galactocentric distances. 
Additionally, mixing of dark matter and GCs caused by violent relaxation during halo assembly can pump up the radial 
extent of a GCS out to an upper limit given by the half-mass radius of the dark matter halo, shown by the dashed line. 
The observed data lie in between these two limits.}
\label{fig:RGCS} 
\end{figure}

\smallskip
In addition to the total mass, we can also place constraints on the radial extent of the \textit{in situ} 
MP GCs by assuming that most MP GCs are formed at $\sim 0.3\Rv$ at $z_{\rm 10\%}$, and remain at the same 
physical distance from the halo center at $z=0$. Indeed, the dynamical friction timescale for a $10^6\msun$ 
GC on a circular orbit at $0.3\Rv$ in a $10^{10}\msun$ halo is $\sim 16\Gyr$, longer than the age of the 
Universe (\citealp{BT08} equation 8.13, assuming a Coloumb logarithm of ${\rm ln}(\Lambda)=10$). 
\Fig{RGCS} shows the results of this simple model in comparison to recent observations by \citet{Forbes17}. The 
observational data show the half-number radii of GCs in a sample of $\sim 30$ GCSs spanning 3 orders of magnitude 
in halo mass, from $M_{\rm 200}\sim 10^{11}-10^{14}\msun$, and including ultra-diffuse galaxies, giant ellipticals 
and the MW. This study found that the radial extent of GCs was proportional to the virial radius of their host halo, 
scaling as $\Mv^{1/3}$. However, we do note that \citet{Hudson17} found a different, non-linear, scaling of the extent 
of GCSs with halo virial radius, albeit using a smaller sample than \citet{Forbes17}. Further observations are needed 
to clarify this point. Comparing our simple model to the observations, we see that while we systematically underpredict 
the radial extent of GCSs by $\sim 0.1-0.3~{\rm dex}$ compared to the observations, the slope of the GCS size-halo mass 
relation is consistent with observations, including the steepening at $\Mv>10^{13}\msun$. In our model, this steepening 
is due to $z_{\rm 10\%}$ decreasing below $z\sim 6$, towards $z\sim 4.8$ and $4$ for $10^{13}$ and $10^{14}\msun$ halos. 

\smallskip
The larger observed radial extents of GCSs compared to our simple model indicates that \textit{ex situ} GCs 
which are accreted onto the halo via mergers tend to lie at larger galactocentric distances than those that 
formed \textit{in situ}. Furthermore, violent relaxation associated with halo assembly from 
$z\sim 6$ to $z=0$ will mix the dark matter and GCs, possibly increasing the effective radius of the GCS. 
The maximal radius that can be reached by this process is the half-mass radius of the dark matter halo. In 
\fig{RGCS} we show the effective radius of an NFW halo as a function of halo mass, assuming the mass-concentration 
relation of \citet{Zhao09}. The slope of this relation is similar to the observed slope, and the observed data 
are roughly midway between this upper limit and our estimated lower limit.

\smallskip
Finally, we stress that our model makes a testable prediction, namely that at high redshift during 
the epoch of GC formation, both MP GCs and less dense star-forming clouds should trace the streams 
that feed galaxies. They should reside outside galaxies, at distances of up to tens of $\kpc$, 
along filamentary, possibly ring like structures. Future observations with JWST are expected to 
observe the formation of GCs \citep{Renzini17,MBK17}, and will be able to test these predictions.

\section{Discussion and Proposed Future Work}
\label{sec:disc}

\smallskip
We have presented a new model for the formation of metal-poor globular clusters (MP GCs), whereby 
they form in the dense streams of gas predicted to feed massive galaxies at high redshift. This 
is reminiscent of older models for the formation of MP GCs in the halos of massive galaxies 
at high redshift \citep{Fall85,Cen01}, but accounts for our new understanding of the filamentary 
nature of the gas in these halos. The filamentary versus spherical structure leads to higher overall 
gas density thus fascilitating gravitational instability. We have shown that such streams are 
unstable to gravitational instabilities on both small and large scales, and that such instabilities 
can collapse in less than a halo crossing time. At high redshift, $z\gsim 4$, the collapsing clouds 
can cool and form stars even if their metalicity is only $Z\sim 0.02 Z_{\odot}$. In the inner $\sim 0.3\Rv$ 
of halos at $z\gsim 6$, the pressure in the streams is large enough to enable the formation of GCs. 
This is already true \textit{prior} to the onset of gravitational instability, and will only be aided 
by gravitationl collapse. The mean densities and pressures in the streams at these redshifts 
are similar to those predicted for the ISM in a Fornax progenitor at $z\sim 3$ \citep{Kruijssen15}, 
and imply a cluster formation efficiency (CFE) of $\gsim 0.3-0.4$ in the streams. This value likely 
increases with halo mass due to increased turbulent pressure. This implies a maximal formation mass 
for GCs of $\sim 6\times 10^5\msun$ in $\sim 10^{10}\msun$ halos at $z\sim 6$, and up to an order of 
magnitude larger in $10^{12}\msun$ halos. Additionally, collisions of counter-rotating streams in the 
inner $0.3\Rv$ can lead to very large compression in the clouds, resulting in enough gas at high enough 
densities to directly form a massive GC.


\smallskip
Our model very naturally accounts for numerous observed properties of MP GCs and GC systems. These include 
the lack of dark matter in GCs, the typical masses of GCs, the correlation between the mass and metalicity 
of MP GCs (as opposed to MR GCs), the correlation between the mass of individual MP GCs and the mass of their 
host halo, the kinematics of MP GCs in the halo, the correlation of the total mass of a MP GC system with the 
mass of its host halo rather than the stellar mass of the central galaxy, and the correlation between the radial 
extent of a GC system and the virial radius of its host halo. Other predictions, such as a spread in formation 
times for MP GCs between $z\sim 4-7$ and a trend such that more massive, more metal rich MP GCs are younger, 
can be tested with future detailed observations that can increase the accuracy of age determinations for MP GCs. 
The key prediction of our model, that the progenitors of MP GCs should trace the streams in the halos around 
galaxies, out to distances of tens of $\kpc$, will be testable with upcoming JWST observations.

\smallskip
Despite the many attractive features of our proposed scenario, we highlight that at the moment the evidence 
is merely suggestive. In \se{model} we have attempted to make quantitative predictions regarding the structure 
and stability of cold streams and the substructure that may form within them, but these are only crude estimates. 
Our model demonstrates that the dense gas in cold streams is likely gravitationally unstable and can lead to 
the formation of stars and stellar clusters in the streams far from the central galaxy. This aspect seems supported 
by cosmological simulations and possibly by recent observations. However, we have not estimated the total expected 
SFR or CFE in the streams, nor have we shown that stream collisions in the inner halo will result in the formation 
of a GC. To properly address the latter, and to better understand the general process of gravitational fragmentation 
of streams in the halo, high resolution numerical simulations will be necessary, as detailed below. Additionally, our 
picture does not account at all for the MR GCs, which likely form inside galaxies \citep{Forbes97,Shapiro10,Kruijssen15} 
rather than in their halos. Our main purpose in this paper was to address the general concept of stream instability at 
high redshift, to highlight a mechanism for the formation of MP GCs directly in the halos of massive galaxies at high 
redshift, and to point to supportive evidence that this scenaro is feasible and can account for many observed properties 
of the MP GC population.

\smallskip
In order to test and refine some of our assumptions and conclusions, future detailed studies of streams will 
be necessary. One aspect of such studies should utilise large scale cosmological simulations to obtain statistics 
regarding the large scale properties of both dark matter filaments and their associated gas streams, such as 
their mass per unit length, radial size, and spin parameter, in much the same way as similar studies have been 
done for dark matter halos and galaxies. By studying such global properties of filaments and how they relate 
to the properties of the dark matter halos they feed, the predictions of \se{model} can be tested and refined. 
Additionally, in a follow-up paper we will examine the frequency and properties of star-forming clumps formed 
in the streams of the full \texttt{VELA} simulation suite.

\smallskip
The second aspect of future studies of cold streams requires detailed studies of the evolution and stability 
of these streams under various conditions. Since the resolution in streams in current cosmological simulations 
is limited \citep{Nelson16}, such studies require idealized simulations of isolated streams inside individual 
halos. A few such studies have been presented recently \citep{M16,Cornuault16,P18}. 
They have shown that streams may be unstable to both hydrodynamical and thermal instabilities, which may lead to 
supersonic turbulence within the streams and to their eventual fragmentation. However, the analysis presented in 
\citet{M16} and \citet{P18}, did not include cooling, the study presented in \citet{Cornuault16} 
included only analytical calculations with no simulations at all and neglected hydrodynamic instabilities, 
and none of these studies accounted for self gravity in the stream or the external gravitational potential of 
the dark matter halo. Additionally, it will be important to directly simulate the collision of counter-rotating 
streams, since this is predicted to be the main mechanism of GC formation in our model. These additional physics 
should be addressed in future work, which will enable the fundamental assumptions and predictions of the model 
presented here to be tested. Until that time, the model presented in this paper may serve as a qualitative guide 
to future studies of the stability of cold streams feeding massive galaxies at high redshift, the origin of MP 
GCs, and the relation between them.

\section*{Acknowledgments}

We thank the anonymous referee for constructive comments which have improved the quality of this manuscript. 
We are greatly indebted to Avishai Dekel for inspiring large parts 
of this work, and for his continuous support. We thank Avishai Dekel and Joel Primack for the use of the 
\texttt{VELA} simulations. We gratefully acknowledge Avishai Dekel, Diederik Kruijssen, and Daisuke Nagai 
for their careful reading of and thoughtful comments on an earlier version of the mauscript. We thank Yuval 
Birnboim, Frederic Bournaud, Andi Burkert, Bruce Elmegreen, Kohei Inayoshi, Thorsten Naab, and Stefanie 
Walch for very helpful discussions. We thank Andrew P. Hearin for his help with the \texttt{Colossus} 
package used in creating \fig{min_mass}. NM and FB acknowledge support from the Klaus Tschira Foundation, 
through the HITS-Yale Program in Astrophysics (HYPA). FB is supported by the US National Science Foundation 
through grant AST-1516962. JB acknowleges support from NSF grant AST-1616598. DC has been funded by the 
ERC Advanced Grant, STARLIGHT: Formation of the First Stars (project number 339177). The \texttt{VELA} 
simulations were performed at the National Energy Research Scientific Computing centre (NERSC), Lawrence 
Berkeley National Laboratory, and at NASA Advanced Supercomputing (NAS) at NASA Ames Research Center (PI: 
Joel Primack).

\bibliography{biblio}

\begin{appendix}

\section{A. Stream Radius and Density in the Case of No Rotation Support}
\label{sec:app}

\smallskip
In \se{model1} we derived expressions for the typical radius and density of cold streams assuming a 
constant ratio of stream radius to the radius of the host dark-matter filament, and that the streams 
were fully rotationally supported. Here we examine the stream radius and density in the case of no 
rotation, assuming purely radial accretion onto the streams from the sheets within which they are 
embedded, and show that the results are inconsistent with both cosmological simulations and observations. 

\smallskip
By combining \equs{flux} and \equss{ML} for the stream line-mass, we obtain one equation relating the stream 
radius and density, 
\be 
\label{eq:rad_dens1}
\Rs^2\rhos \simeq 6.1\times 10^6 \msun \kpc^{-1}~ \M10^{0.77} \z7^2 \fst \Machv^{-1}.
\ee
{\no}A second equation can be obtained by considering the case of radial accretion onto the stream from 
a background with density $\rho_{\rm acc}$ flowing at velocity $V_{\rm acc}$ towards the stream axis. 
The accretion rate onto the stream in this case can be written as 
\be 
\label{eq:accretion_rate_toy}
{\dot {M}}_{\rm L,s} = 2\pi \Rs V_{\rm acc} \rho_{\rm acc}.
\ee
{\no}The accretion rate onto the stream can be obtained by taking the derivative of \equ{ML} with respect 
to time, and inserting \equ{Mdotv} as well as $\z7\simeq (t/0.95~\Gyr)^{-2/3}$, and is given by \equ{stream_accretion}. 
The accretion velocity is given by the gravitational potential of the dark-matter filament and the gas stream, and is 
given by \equ{accretion}. To within a factor of $\lsim 2$, this is the virial velocity of the dark-matter filament. 

\smallskip
What remains is to relate the density in the stream, $\rhos$, to that of the accreted material, $\rho_{\rm acc}$. 
Since the cooling time in the sheets is short, the accreted gas is isothermal \citep{Mo05} and we expect a strong 
accretion shock at the stream boundary where the density increases by a factor $\mathcal{M}_{\rm acc}^2$, where 
\be 
\label{eq:accretion_mach_radial}
\mathcal{M}_{\rm acc} = V_{\rm acc}/\cs \simeq 7.8~ \M10^{0.38} \z7 \fst^{0.5} \Machv^{-0.5} \T4^{-0.5}.
\ee
{\no}is the Mach number of the accreted material. We thus have $\rhos \simeq \mathcal{M}_{\rm acc}^2 \rho_{\rm acc}$. 
Putting all this together, we obtain a second equation relating the stream radius and density: 
\be 
\label{eq:rad_dens2}
\Rs \rhos \simeq 7.9\times 10^6 \msun \kpc^{-2}~ \M10^{1.15} \z7^{4.5} \fst^{1.5} \Machv^{-1.5} \T4^{-1}.
\ee

\smallskip
Combining \equs{rad_dens1} and \equss{rad_dens2} we obtain 
\be 
\label{eq:radial_Rs}
\Rs \simeq 0.8 \kpc~ \M10^{-0.38} \z7^{-2.5} \fst^{-0.5} \Machv^{2.5} \T4,
\ee
{\no}and
\be 
\label{eq:radial_Ds}
n_{\rm H,s} \simeq 0.3 \cmc~ \M10^{1.53} \z7^7 \fst^2 \Machv^{-4} \T4^{-2}.
\ee
{\no}Note that these are the values at $\Rv$, before the stream enters the halo. Within the halo, 
the radius scales as $\xv=R/\Rv$, while the density scales as $\xv^{-2}$. 

\smallskip
Somewhat amazingly, the above estimate assuming purely radial accretion and the estimate from \se{model1} 
assuming total rotational support both agree for a $10^{10}\msun$ halo at $z=6$. However, when scaled to 
a $10^{12}\msun$ halo at $z=3$, the above estimate yields densities of $n_{\rm H,s}\sim 6.85\cmc$, several 
orders of magnitude larger than the densities seen in cosmological simulations \citep[e.g.][]{Goerdt10,FG10,
vdv12}, and inferred from observations of giant Ly$\alpha$ nebulae around massive galaxies at $z\sim 3$ 
\citep[e.g][]{Cantalupo14,Arrigoni15,Borisova16,Ginolfi17}. Furthermore, for the same halo mass and redshift 
\equ{radial_Rs} results in a ratio of stream radius to halo virial radius of $\Rs/\Rv\sim 7\times 10^{-3}$, 
significantly smaller than what is seen in simulations. Indeed, such small streams are predicted to rapidly 
disrupt in the halo due to Kelvin-Helmholz Instabilities (\citealp{M16}; Padnos et al., in preparation).

\smallskip
We conclude that the streams are not fed by purely radial accretion from their surroundings, and that the 
accreted material must have some rotational support. This is consistent with the large vorticities present 
in the stream gas \citep{Codis12,Codis15,Laigle15}, and justifies our derivation of the stream radius and 
density assuming rotational support in \se{model1}.

\section{B. Effect of a Tidal Field on the Fragmentation Scale}
\label{sec:app2}

\smallskip
In \se{model2} we derived the characteristic fragmentation scale in the streams using the Jeans stability 
criterion. However, in the presence of a tidal field these scales may be modified. The modified Jeans length 
can be approximated as \citep{Jog13} 
\be 
\label{eq:tidal}
\lambda'_{\rm j} = \frac{\Lj}{\left[1-T_0/(4\pi G \rho_0)\right]^{1/2}}, 
\ee
{\no}where $\Lj$ is the Jeans length in the absence of the tidal field (\equ{LJeans_def}), $\rho_0$ is the 
unperturbed background density, and $T_0=\partial^2 \Phi / \partial r^2$ is the second derivative of the 
gravitational potential, estimated at the center of the region. If $T_0>0$ then the tidal field is extensive 
and the fragmentation scale length is larger than in the case with no tidal field. If $T_0<0$ the tidal field 
is compressive and the fragmentation scale decreases. There are two sources of tidal forces on a cloud in the 
center of the streams: the dark matter halo and the dark matter filament. We estimate the ratio $T_0/(4\pi G \rhos)$ 
for each of these. 

\smallskip
For the halo, we assume it is an isothermal sphere with $M(r)\propto r$, resulting in $T_{\rm 0,halo}=(\Vv/\Rv)^2\xv^{-2}$. 
Together with \equs{Rvir}, \equss{Vvir} and \equss{Rhos} this yields 
\be 
\label{eq:tidal_sphere}
\frac{T_{\rm 0,halo}}{4\pi G \rhos} \simeq 0.08\L1^2.
\ee
{\no}This is independent of halo mass or redshift. For $\L1=1$, this results in a $\sim 4\%$ increase in the Jeans 
length relative to \equ{LJeans}, and a corresponding $\sim 12\%$ increase in the Jeans mass relative to \equ{Mclump}. 
For $\L1=0.5$, as was often assumed in our analysis, the difference is even smaller. This is negligible relative to 
the other uncertainties in our model.

\smallskip
For the dark matter filament, we assume that it is an isothermal filament with a line-mass profile given by \equ{Mlr}. 
This results in $T_{\rm 0,Fil}=-\pi G \rho_{\rm dm,c}$, where $\rho_{\rm dm,c}$ is the central density of the dark matter 
filament. This is related to the total line-mass by $\rho_{\rm dm,c} = \Mlf/(\pi r_{\rm h}^2)$, with 
$r_{\rm h}=\beta^{-1} r_{\rm v,Fil}$ the half-mass radius of the filament. Note that since $T_{\rm 0,Fil}<0$, 
the tidal field is compressive. Putting all this together yields 
\be 
\label{eq:tidal_fil}
\frac{\left| T_{\rm 0,Fil}\right|}{4\pi G \rhos} \simeq \frac{0.014\L1^2}{\beta^2}.
\ee
{\no}For any $\beta>1$, the effect on the fragmentation scale is completely negligible.

\end{appendix}

\end{document}